\documentclass[useAMS,usenatbib]{mn2e}
\usepackage{graphicx}
\usepackage{times}
\usepackage{epsfig}
\usepackage{longtable}
\usepackage{lscape}
\usepackage{amssymb}
\usepackage{latexsym}
\usepackage{natbib}
\usepackage{afterpage}
\def\kms{$\rm km\;s^{-1}$}
\def\deg{^{\circ}}

\title[Signatures of accretion events in galaxy halos]{Signatures of accretion events in the halos of early-type galaxies from comparing PNe and GCs kinematics}
\author[L. Coccato et al.]{L. Coccato$^{1}$\thanks{E-mail: lcoccato@eso.org}, M. Arnaboldi$^{1,2}$, O. Gerhard$^{3}$
\\
$^{1}$European Southern Observatory, Karl-Schwarzschild-Stra$\beta$e 2, 85748 Garching bei M\"unchen, Germany.\\
$^{2}$INAF, Osservatorio Astronomico di Pino Torinese, I-10025 Pino Torinese, Italy.\\
$^{3}$Max-Planck-Institut f\"ur extraterrestrische Physik, Postfach 1312, Giessenbachstr., D-85741 Garching bei M\"unchen, Germany.}
\begin{document}

\date{received: xxx; accepted: yyy}

\pagerange{\pageref{firstpage}--\pageref{lastpage}} \pubyear{2012}

\maketitle

\label{firstpage}

\begin{abstract}
  We have compared the halo kinematics traced by globular clusters
  (GCs) and planetary nebulae (PNe) for two elliptical galaxies
    in the Fornax and Virgo clusters NGC 1399 and NGC 4649, and for
  the merger remnant NGC 5128 (Centaurus A). We find differences
  in the rotational properties of the PN, red GC, and blue GC
  systems in all these three galaxies. 
 NGC 1399 PNe and GCs show line of sight velocity distributions in
  specific regions that are significantly different, based on
  Kolmogorov-Smirnov tests. The PN system shows multi-spin components,
  with nearly opposite direction of rotation in the inner and the
  outer parts. The GCs velocity field is not point-symmetric in the
  outer regions of the galaxy, indicating that the system has not
  reached dynamical equilibrium yet. In NGC 4649 PNe, red and blue GCs
  have different rotation axes and rotational velocities. Finally, in
  NGC 5128 both PNe and GCs deviate from equilibrium in the outer
  regions of the galaxy, and in the inner regions the PN system is
  rotationally supported, whereas the GC system is dominated by
  velocity dispersion.
The observed different kinematic properties, including deviations from
point-symmetry, between PNe and GCs suggest that these systems are
accreted at different times by the host galaxy, and the most recent
accretion took place only few Gyr ago. We discuss two scenarios which
may explain some of these differences: i) tidal stripping of
loosely-bound GCs, and ii) multiple accretion of low luminosity and
dwarf galaxies. Because these two mechanisms affect mostly the GC
system, differences with the PNe kinematics can be expected.

\end{abstract}


\section{Introduction}
  
Stellar halos of galaxies offer an important laboratory to understand
 the processes of galaxy formation and evolution. The dynamic
time scales in the halos are large\footnote{They are of the order of
  1 Gyr (estimated for distance $R = 50$ kpc and circular velocity
  $V_C = 250$ \kms, e.g. \citealt{Binney+87}), which represents a
  significant fraction of the age of the universe.}, and the imprint
of the formation mechanisms may still be preserved at large radii in
the kinematics and orbital structure, in streams and
substructures, or in the chemical composition and distribution of
stars.
  Therefore, if the last interaction or merger episodes that build up
  the galaxy halo occurred only few Gyr ago, the halo may not have
  reached dynamical equilibrium yet. For the case of elliptical
    galaxies, we can test the dynamic state of the halo by simple
  symmetry arguments. Elliptical galaxies are intrinsically triaxial
  systems (e.g. \citealt{Statler94} and references therein), hence
  their projected light and kinematics on the sky must be consistent
  with point symmetry when in equilibrium. We can quantify
  perturbations from this status by measuring deviations from such
  point symmetry as a function of radius. This simple approach does not
  require the full computation of the equilibrium status, and measures
  a basic property of the kinematical quantities projected on
  the sky.

Due to the rapid dimming of the stellar light at large radii,
  discrete tracers of the underlying stellar population must be
  used. Planetary Nebulae (PNe) and Globular Clusters (GCs) can be
detected in large numbers out to several effective radii. They and have
been used to probe the kinematics \citep[e.g.][]{Coccato+09,
  Herrmann+09, Teodorescu+10, Lee+10, Schuberth+12, Pota+13}, dark
matter content \citep[e.g.][]{Cote+01, Romanowsky+03, deLorenzi+09,
  Napolitano+11, Richtler+11, Deason+12} and chemical composition
\citep[e.g.][]{Forbes+11, Alves+11} of the stellar halos in nearby
galaxies, probing regions much further out than those reachable
  with absorption-line kinematics (e.g. \citealt{Weijmans+09,
    Coccato+10a, Greene+12}).
PNe have been found to be good tracers of the underlying stellar
population, as their spatial distribution and kinematics agree well
with those of the stars \citep{Coccato+09, Cortesi+13a}. This is {\bf
  not self evident} for GCs: they have a bi-modal color distribution
(see the review by \citealt{BrodieStrader06} and references therein)
where red GCs have a radial density profile similar to that of the
stars and PNe, while blue GCs have a more extended spatial
distribution (e.g. \citealt{Pota+13, Schuberth+10}).

\begin{figure*}
%~/data/halos/kinematics/paper/phase.space.figures/mkfigure.pro   
%;data from repository/object_separation<galaxy>/<component>/<histogram>/hist<galaxy>.pro
%\hspace{-.3cm}
%\vbox{
% \hbox{
%  \psfig{file=phase.1399.ps,width=7.6cm,clip=}   %7.2
%  \psfig{file=combined.prob.pne.1399.ps,width=5.cm,clip=}%5.0
%  \psfig{file=combined.prob.gcs.1399.ps,width=5.cm,clip=}}
% \hbox{
%  \psfig{file=phase.4649.ps,width=7.6cm,clip=}
%  \psfig{file=combined.prob.pne.4649.ps,width=5.cm,clip=}
%  \psfig{file=combined.prob.gcs.4649.ps,width=5.cm,clip=}}
%}
\vbox{
 \hbox{
  \psfig{file=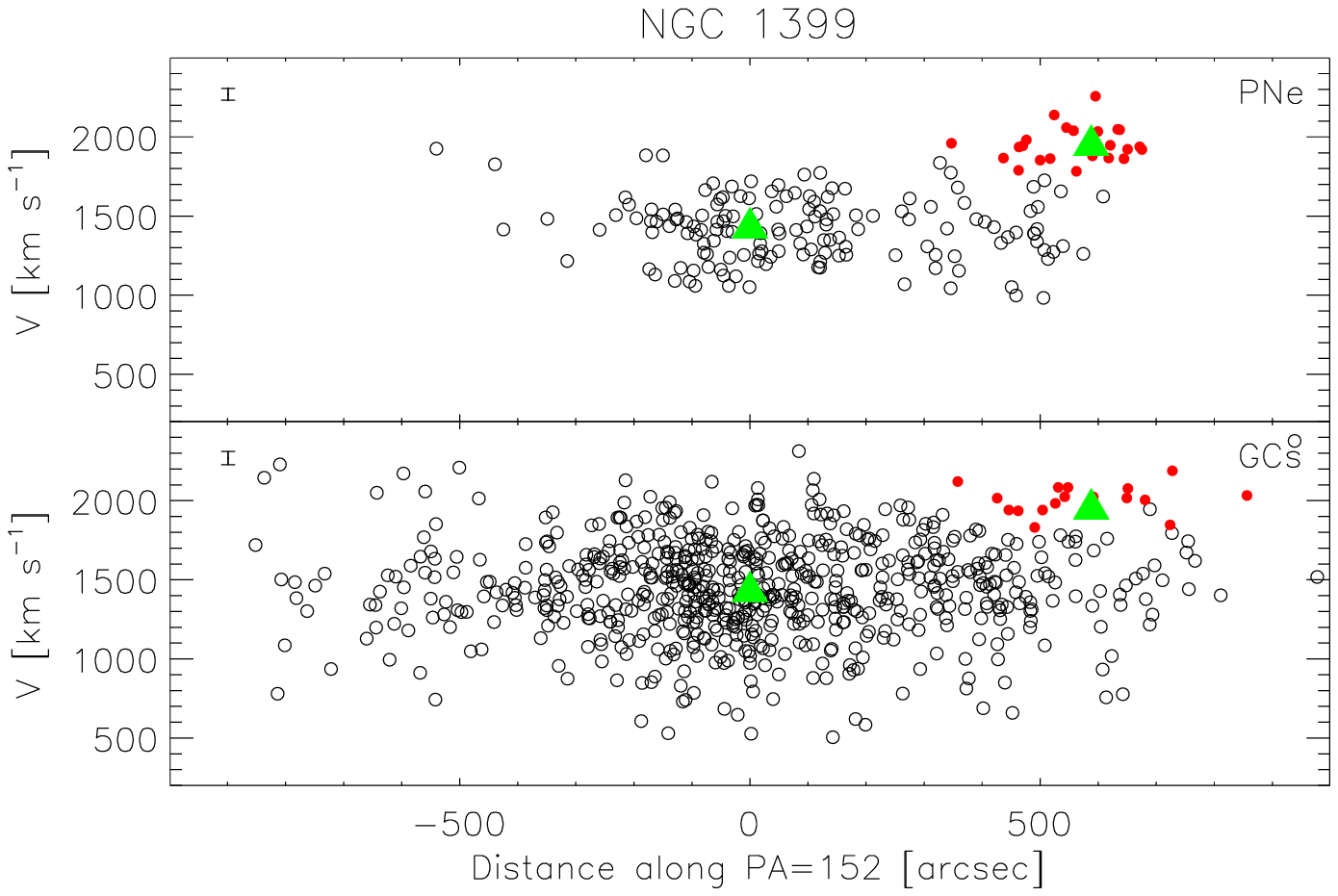,width=8.9cm,clip=}   %7.2
  \psfig{file=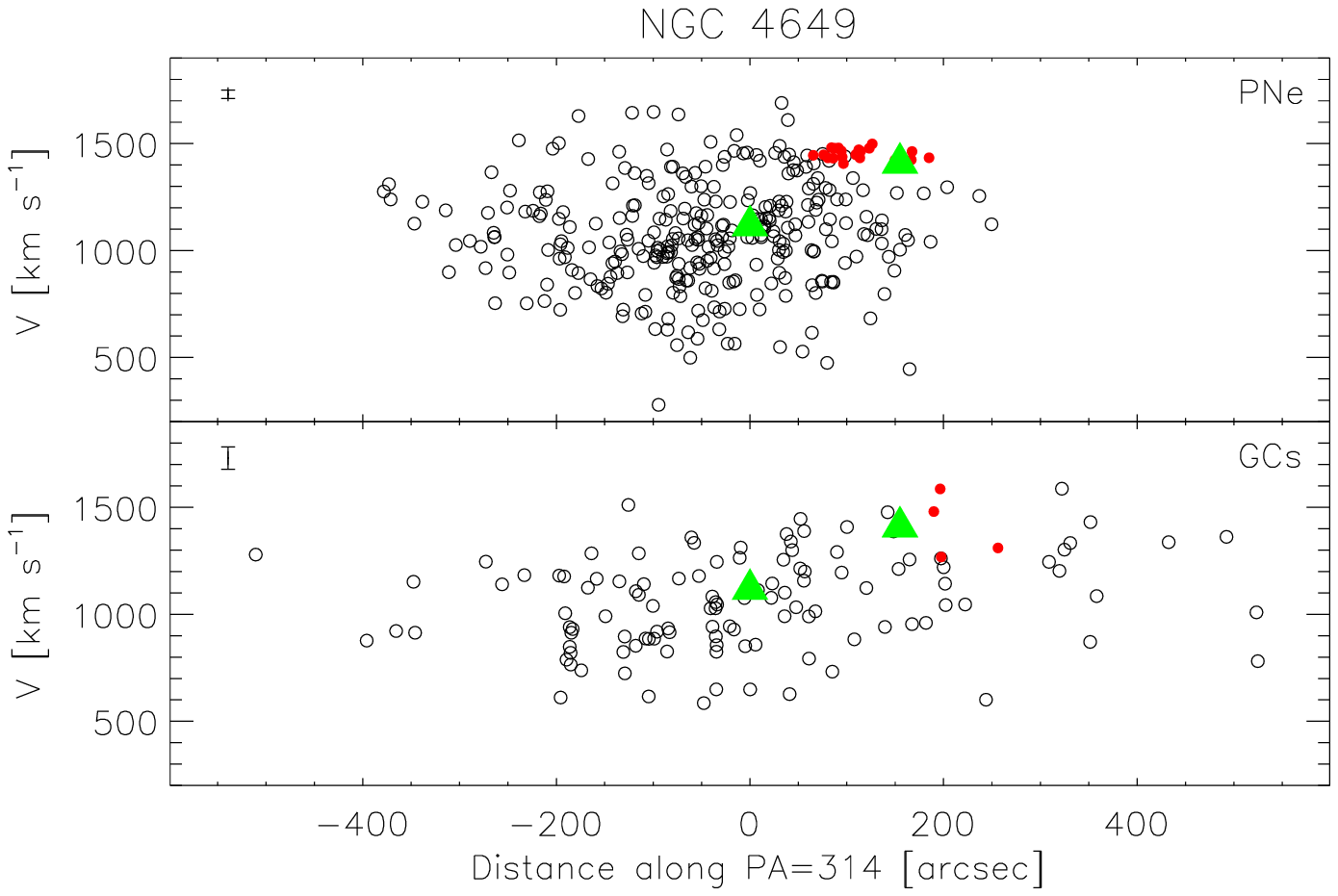,width=8.9cm,clip=}
 }
 \hbox{
  \psfig{file=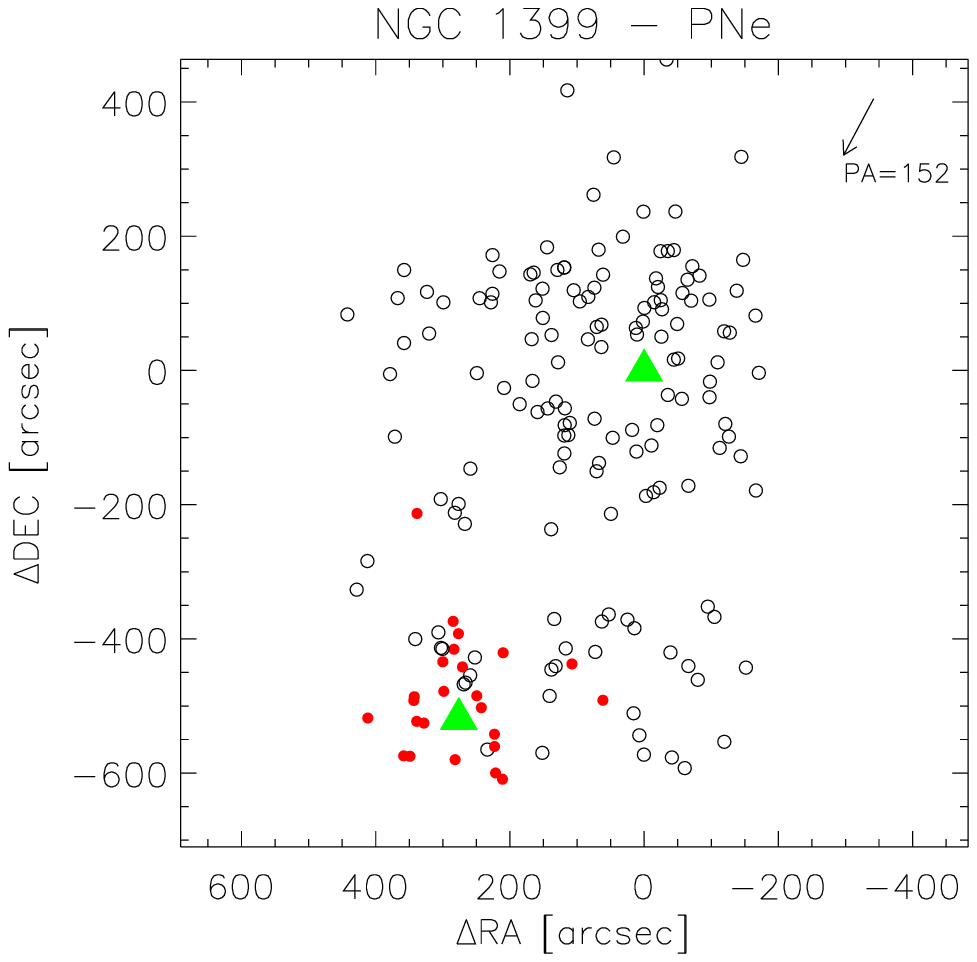,width=7.9cm,clip=}%5.0
  \psfig{file=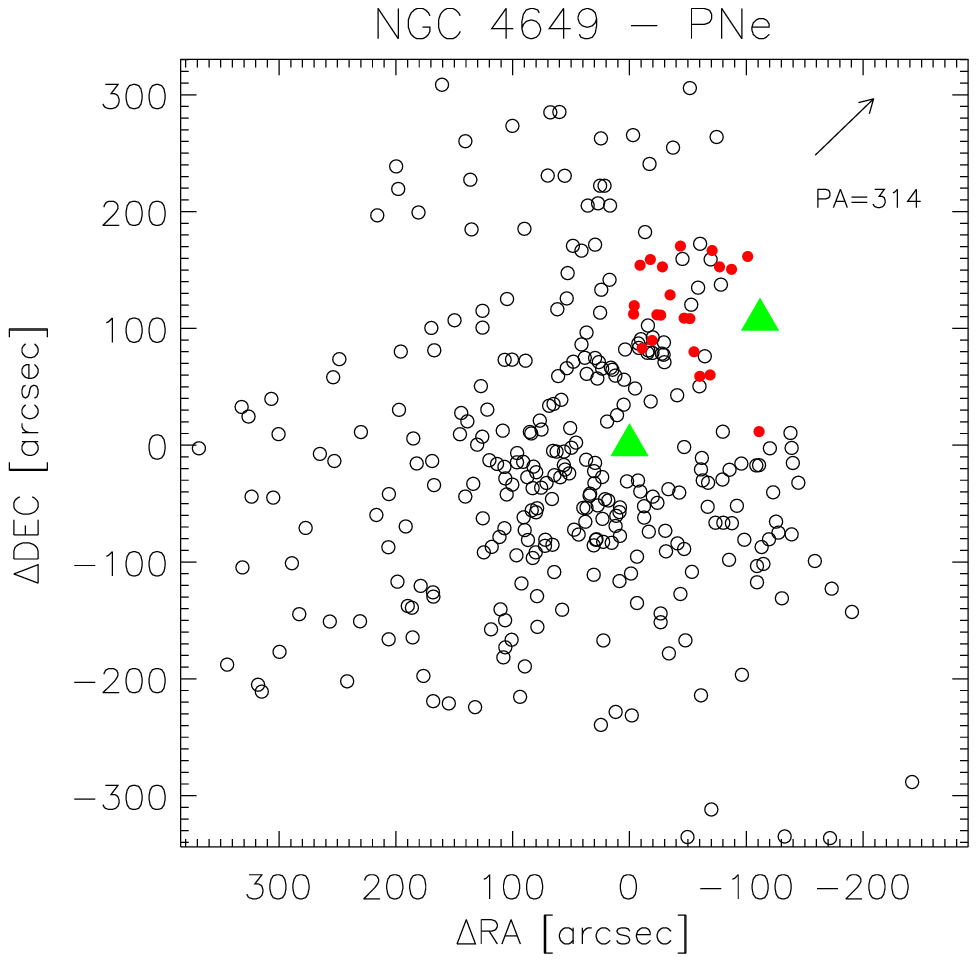,width=7.9cm,clip=}
 }
 \hbox{
  \psfig{file=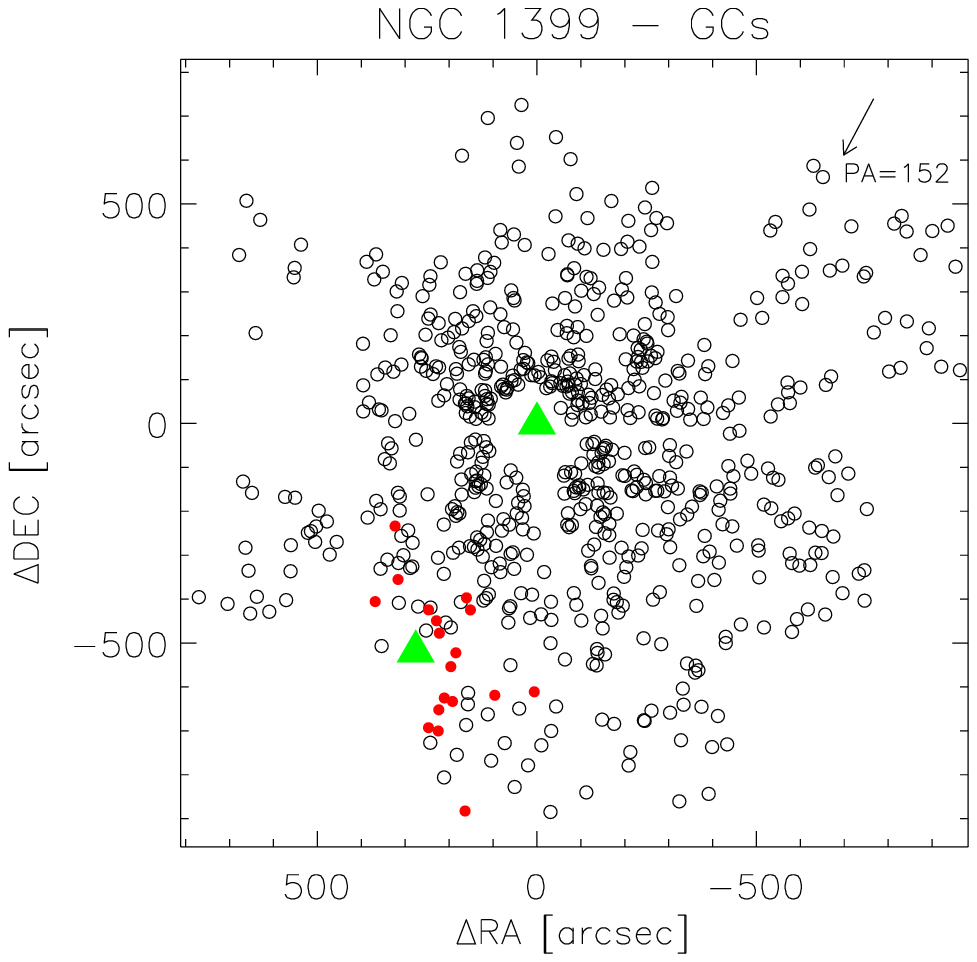,width=7.9cm,clip=}
  \psfig{file=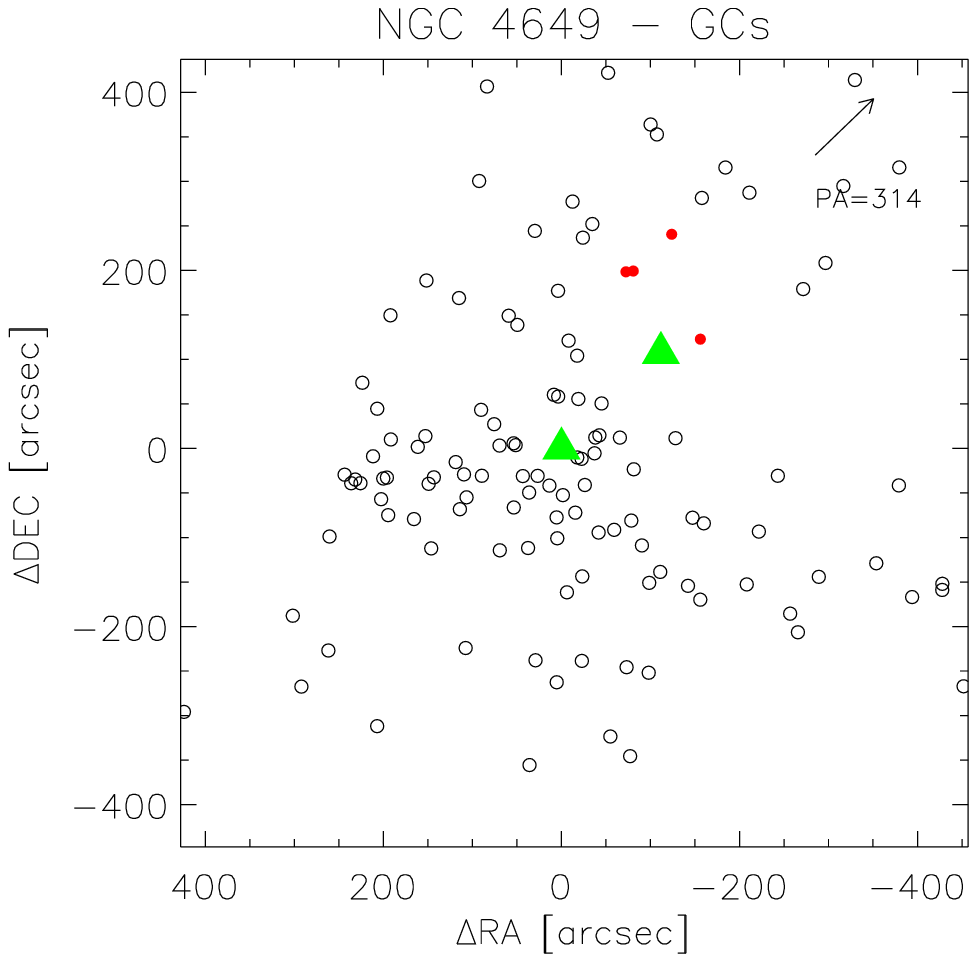,width=7.9cm,clip=}
 }
}
\caption{Membership allocation of PNe and GCs between the main
  galaxies (NGC~1399 and NGC~4649) and their companions (NGC~1404 and
  NGC~4647). The top panels show the radial velocity of PNe and
  GCs as a function of their position along the line connecting the
  main galaxy to the companion. The vertical bar on the top-left
    corner of each plot indicates the mean error on the measured
    radial velocity. The  middle and bottom panels show the
  location of the PNe and GCs on the sky. Red filled circles:
    tracers that have more than 60\% probability to belong to the
  satellite, and therefore are removed from the analysis. Black
    open circles: tracers that have less than 60\% probability to
    belong to the satellite. Green triangles mark the position of
  galaxies and satellites.  The arrows on the top-right corner
  indicate the direction along which the position velocity diagrams in
  the left panels are generated. North is up, East is left.}
\label{fig:membership}
\end{figure*}

\begin{figure*}
% ~/data/halos/kinematicskinematics/README - idl -e "execute_kernel_smoothing,....
\hspace{.5cm}
\vbox{
  \hbox{
    \psfig{file=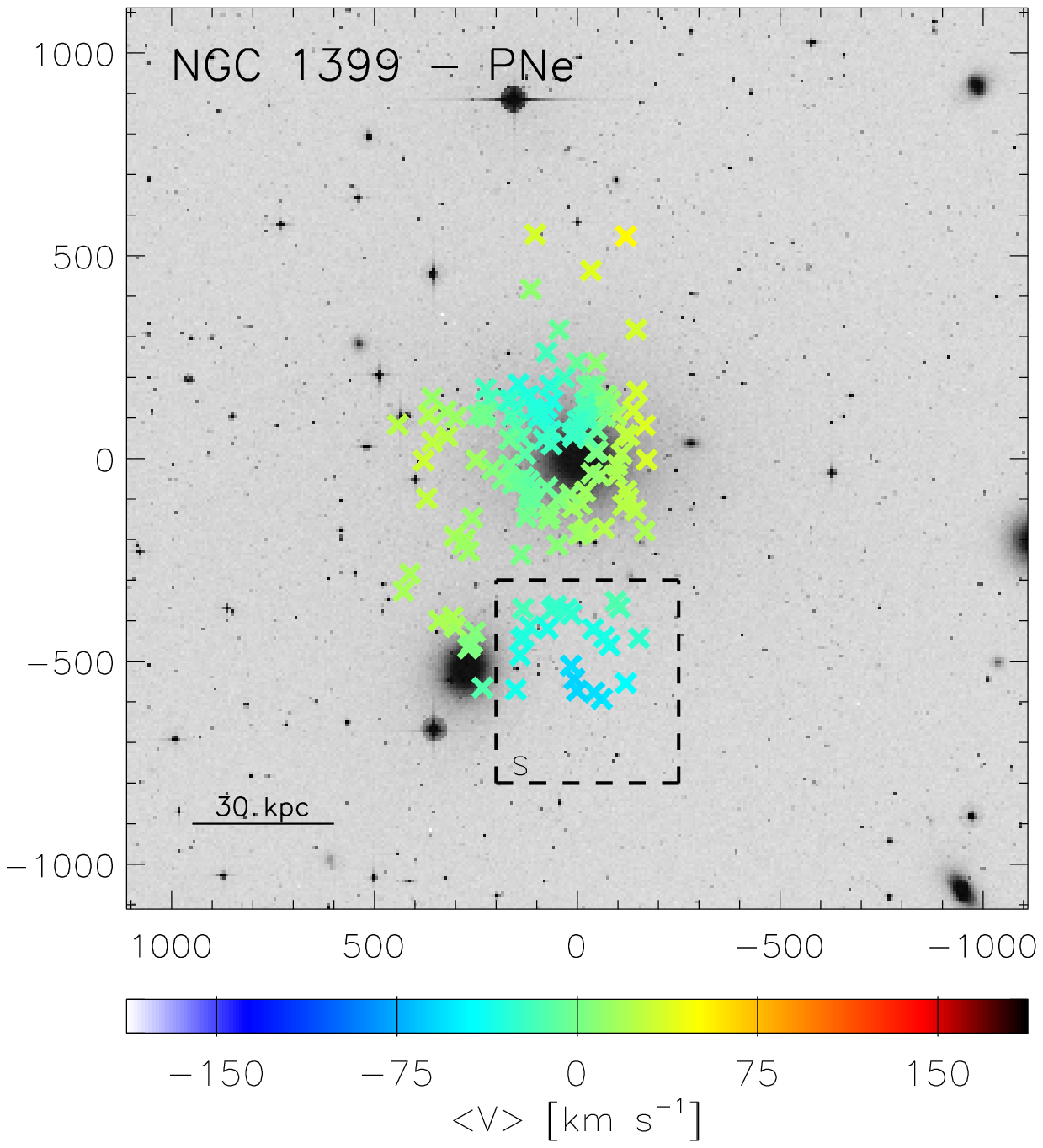,width=5.2cm,clip=,bb=53 400 404 733}   %5.2cm
    \psfig{file=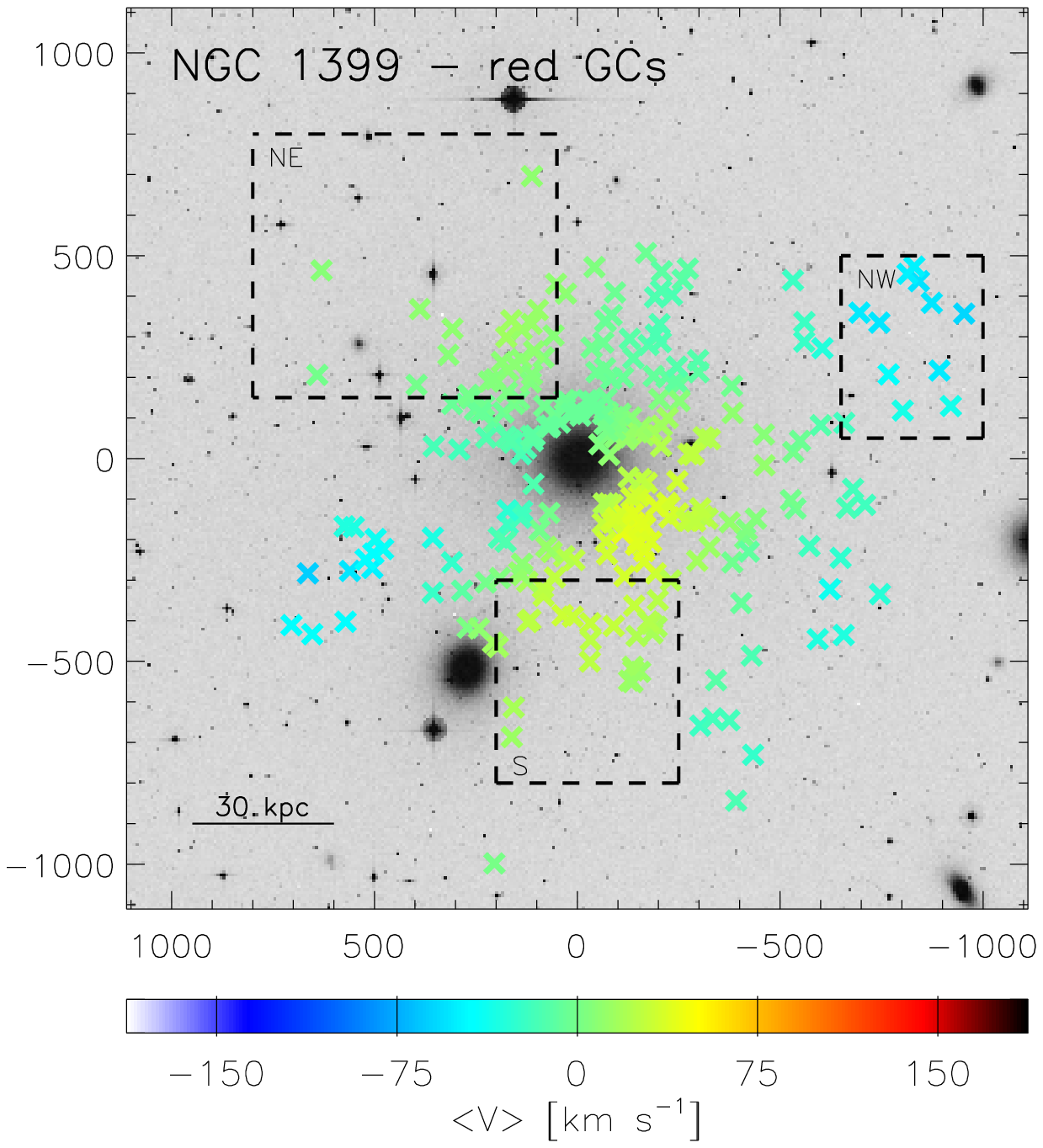,width=5.2cm,clip=,bb=53 400 404 733}
    \psfig{file=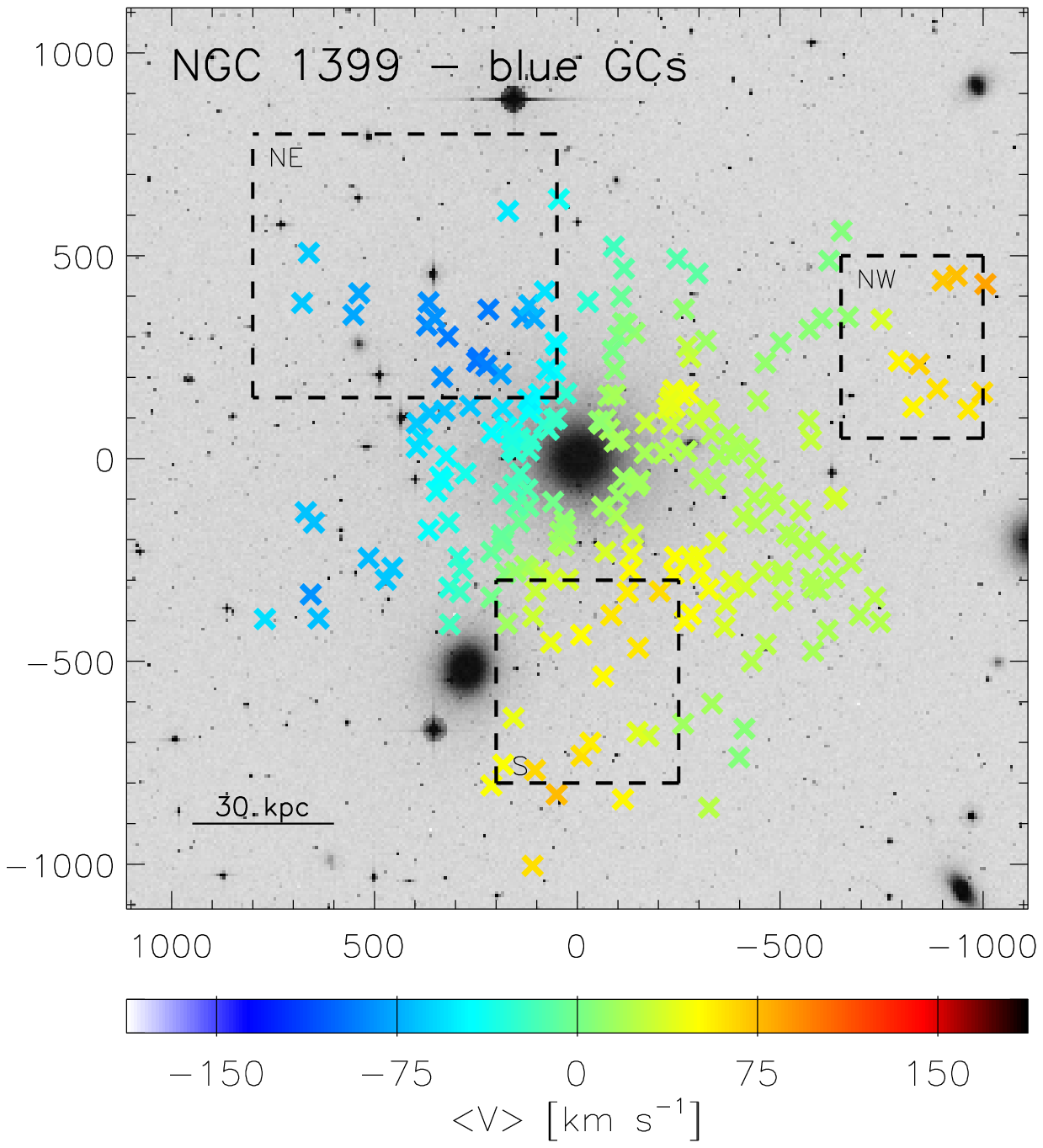,width=5.2cm,clip=,bb=53 400 404 733}
     \psfig{file=1399.pne.ps,width=4.8cm,clip=,bb=74 330 404 400,angle=90}}%4.8cm ; 4.6
  \hbox{
    \psfig{file=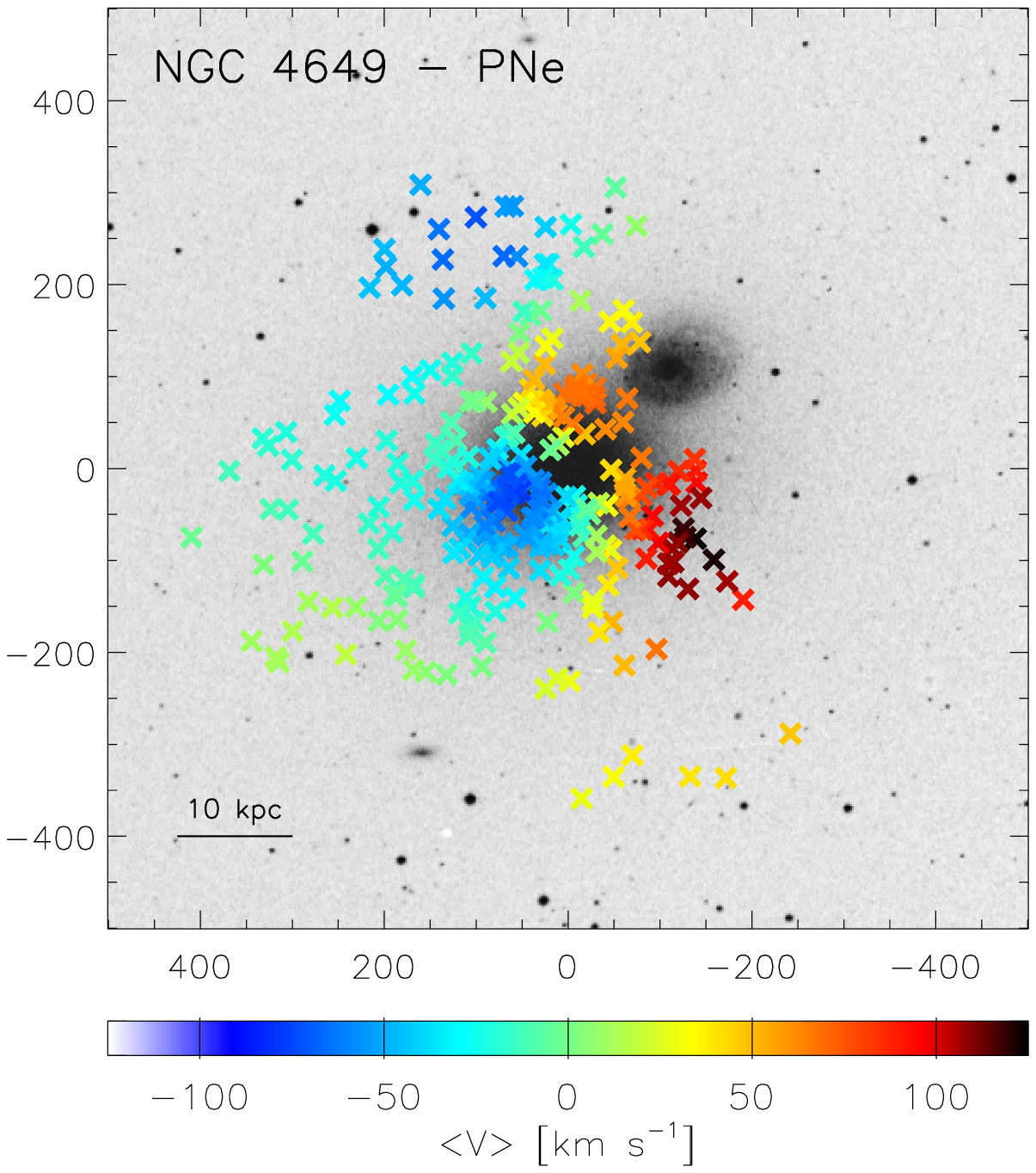,width=5.2cm,clip=,bb=53 400 404 733}   %5.2cm
    \psfig{file=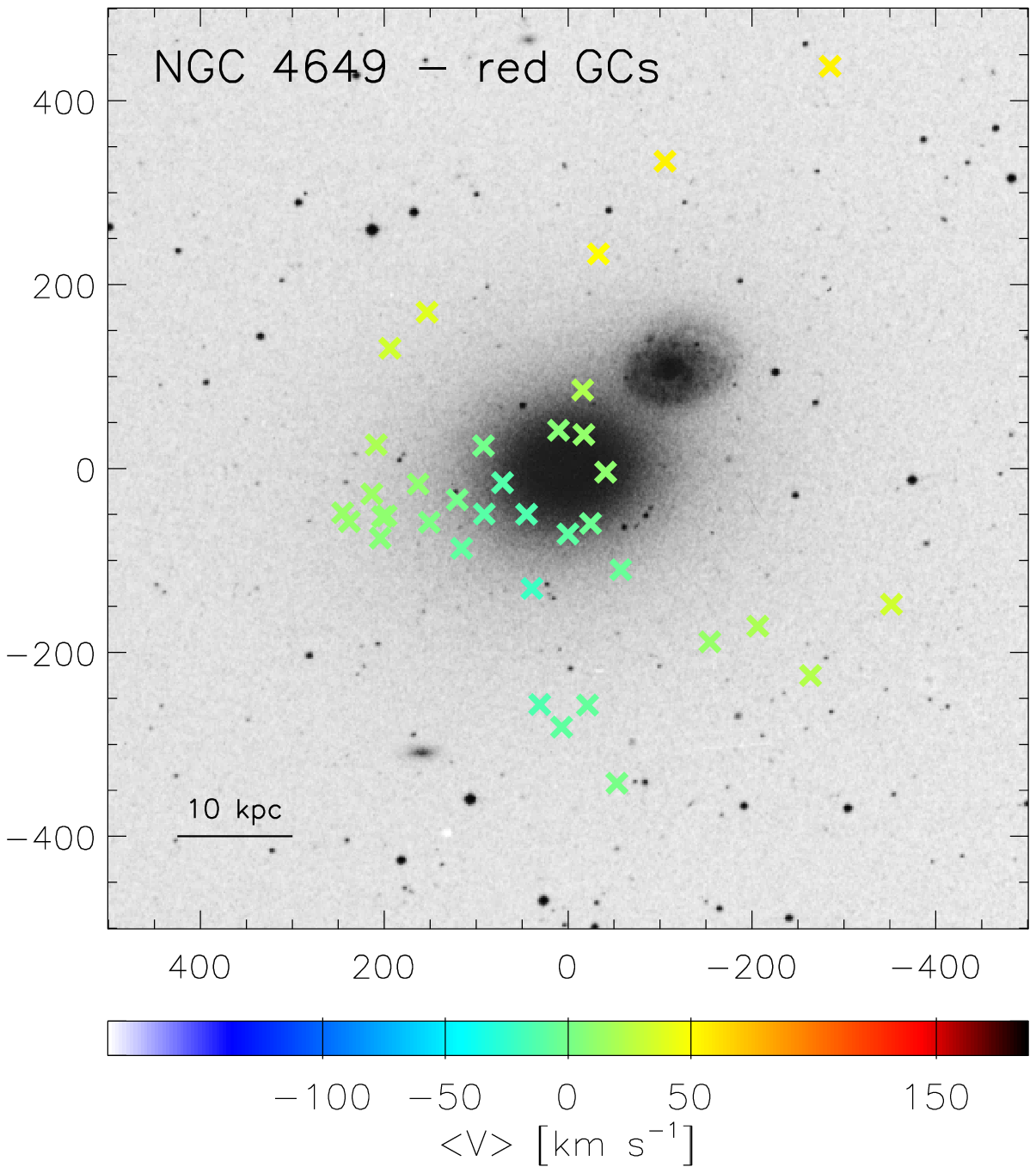,width=5.2cm,clip=,bb=53 400 404 733}
    \psfig{file=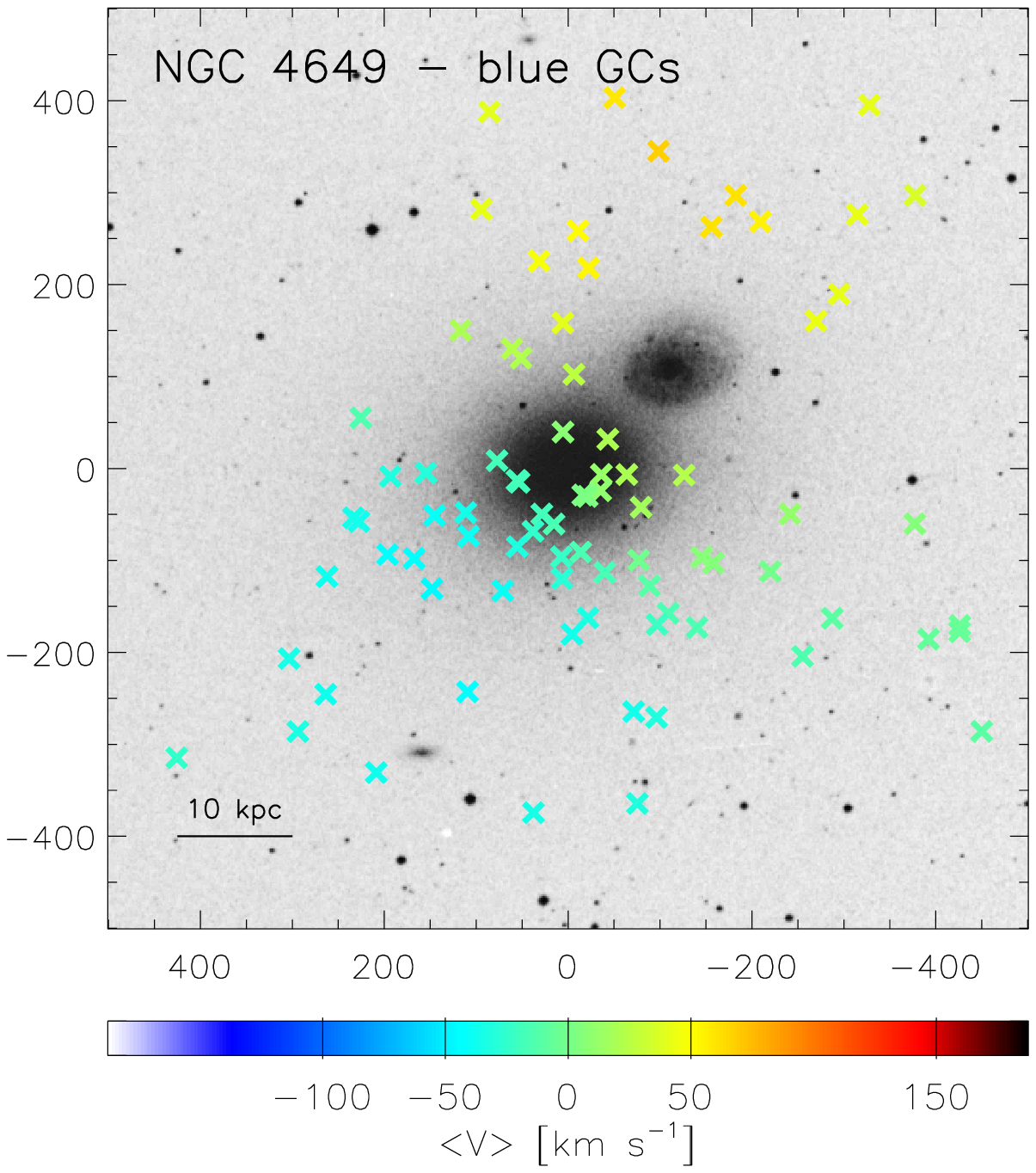,width=5.2cm,clip=,bb=53 400 404 733}
     \psfig{file=4649.pne.ps,width=4.8cm,clip=,bb=74 330 404 400,angle=90}}%4.8cm ; 4.6
  \hbox{
    \psfig{file=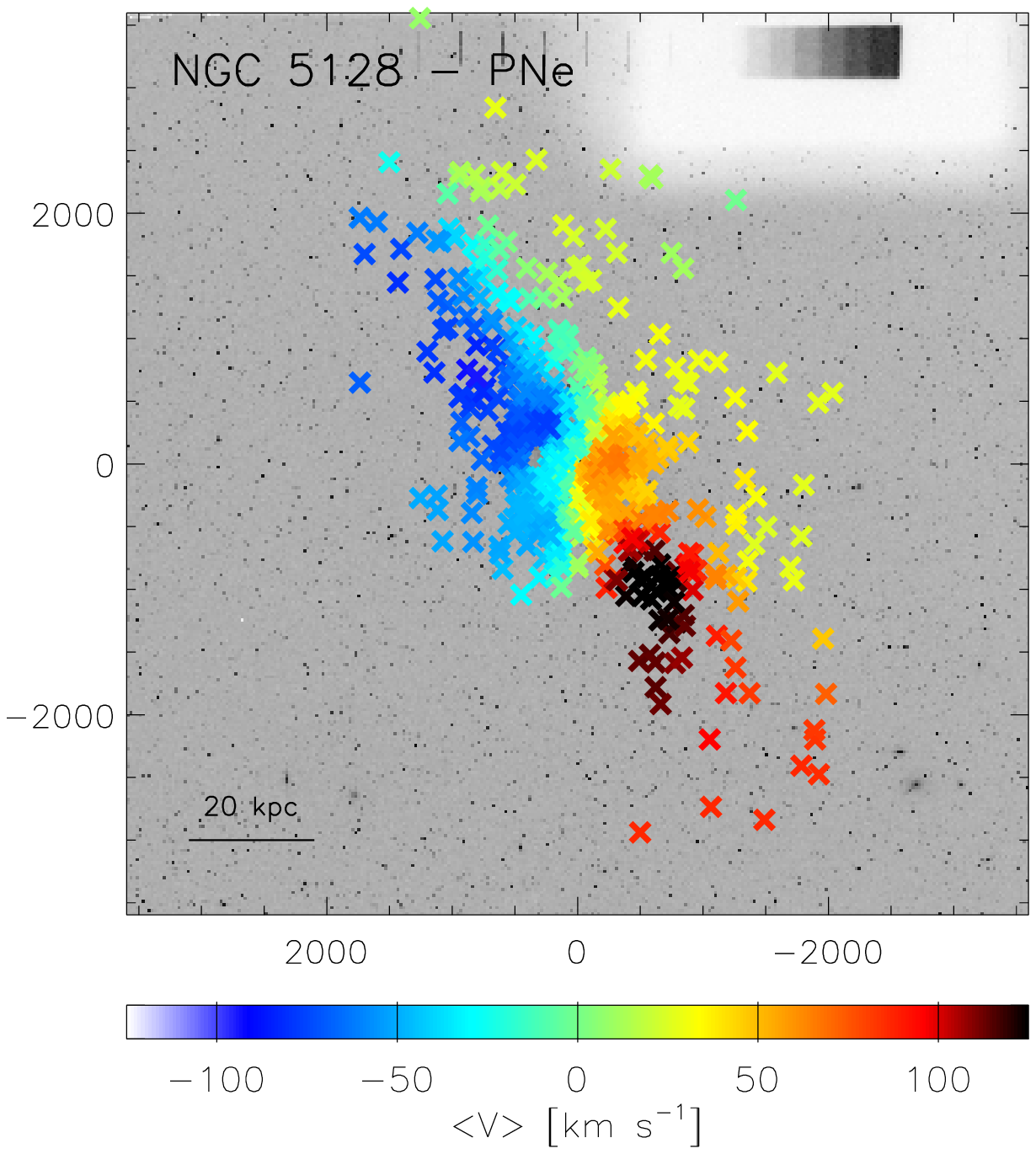,width=5.2cm,clip=,bb=53 400 404 733}   %5.2cm
    \psfig{file=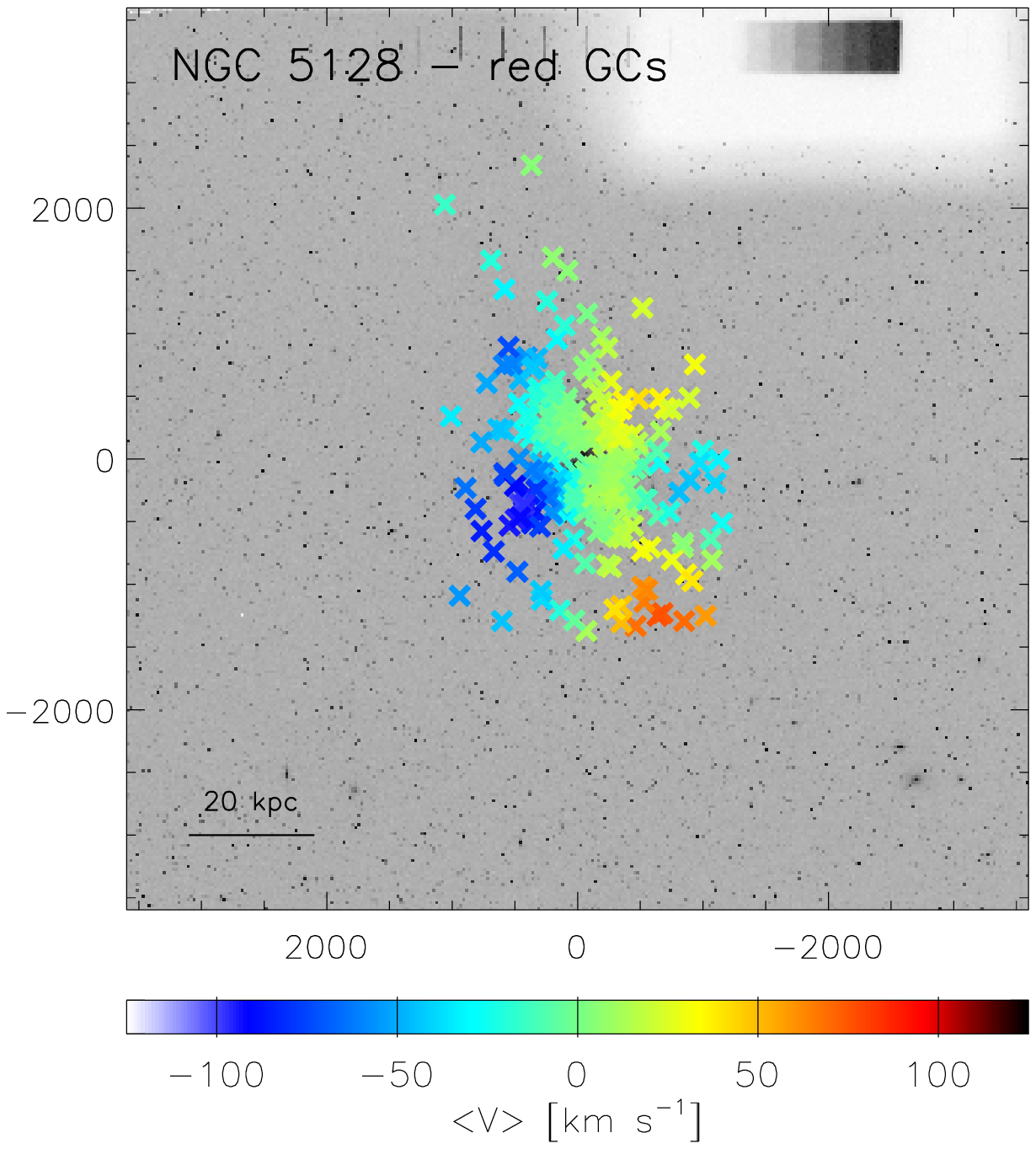,width=5.2cm,clip=,bb=53 400 404 733}
    \psfig{file=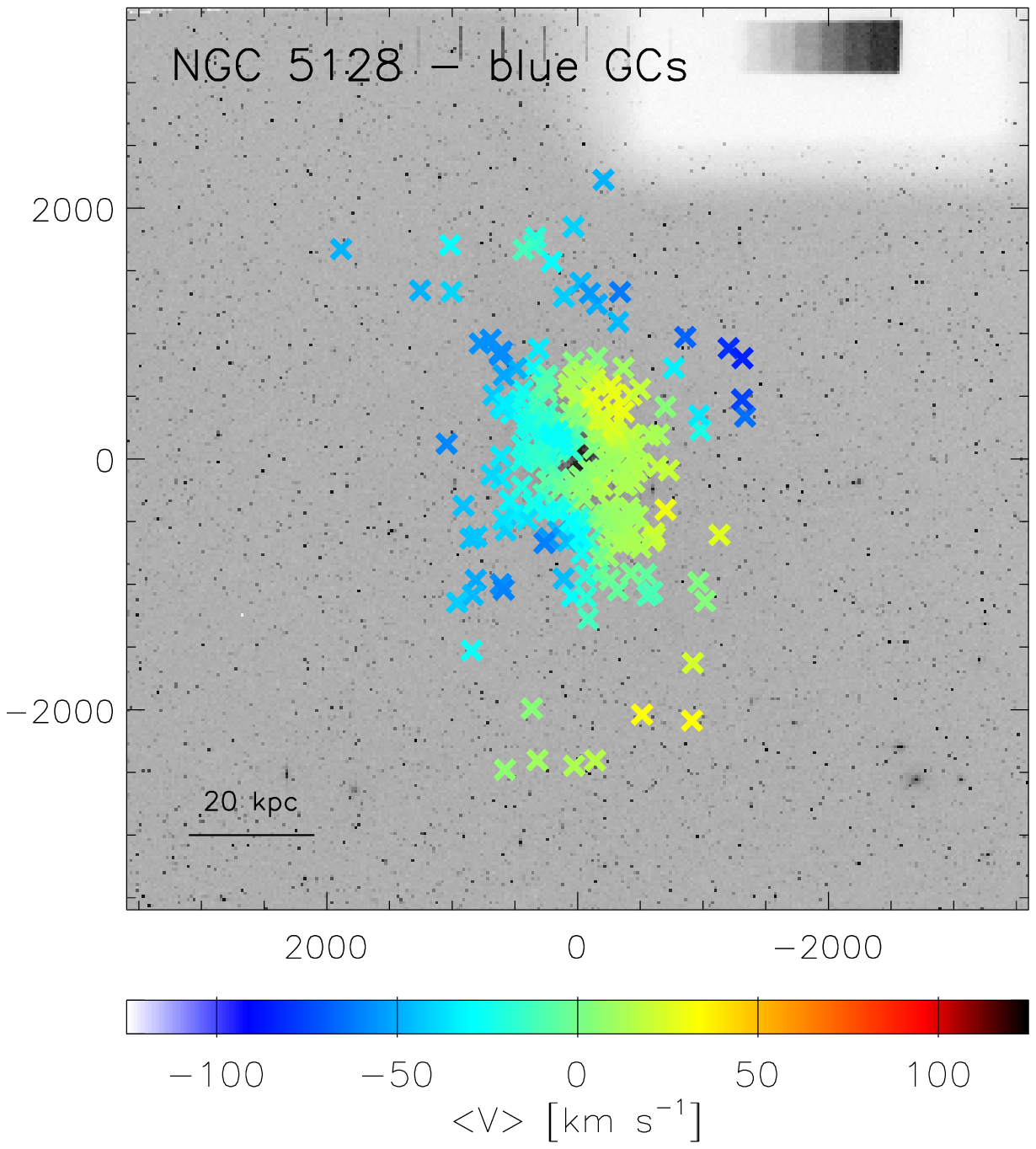,width=5.2cm,clip=,bb=53 400 404 733}
    \psfig{file=5128.pne.ps,width=4.8cm,clip=,bb=74 330 404 400,angle=90}}%4.8cm ; 4.6
%   \psfig{file=1399.red.ps,width=5.6cm,clip=,bb=53 400 404 733}
%   \psfig{file=4649.red.ps,width=5.6cm,clip=,bb=53 400 404 733}
%   \psfig{file=5128.red.ps,width=5.6cm,clip=,bb=53 400 404 733}}
% \hbox{
%   \psfig{file=1399.blue.ps,width=5.6cm,clip=,bb=53 330 404 733}
%   \psfig{file=4649.blue.ps,width=5.6cm,clip=,bb=53 330 404 733}
%   \psfig{file=5128.blue.ps,width=5.6cm,clip=,bb=53 330 404 733}}
  }
  \caption{Mean velocity fields for NGC 1399 (top panels), NGC 4649
    (middle panels) and NGC 5128 (bottom panels), with the galaxy
    systemic velocity subtracted. Each panel shows the galaxy image,
    where the positions of the PNe and GCs are marked with crosses:
    from left to right the panels refers to the PNe, red GC, and blue
    GC sub-populations, respectively. The colors indicate the values
    of the mean velocity field $\langle V \rangle$ computed at that
    position. North is up, East is left. Coordinates are in
    arcseconds. The dashed lines define the regions where the
    line-of-sight velocity distributions of PNe and GCs kinematics are
    compared locally (see text for details).}
\label{fig:fields} 
\end{figure*}

%\clearpage 

 \begin{figure*}
 % ~/data/halos/kinematicskinematics/README - idl -e "execute_kernel_smoothing,....
 \hspace{.5cm}
 \vbox{
  \hbox{
     \psfig{file=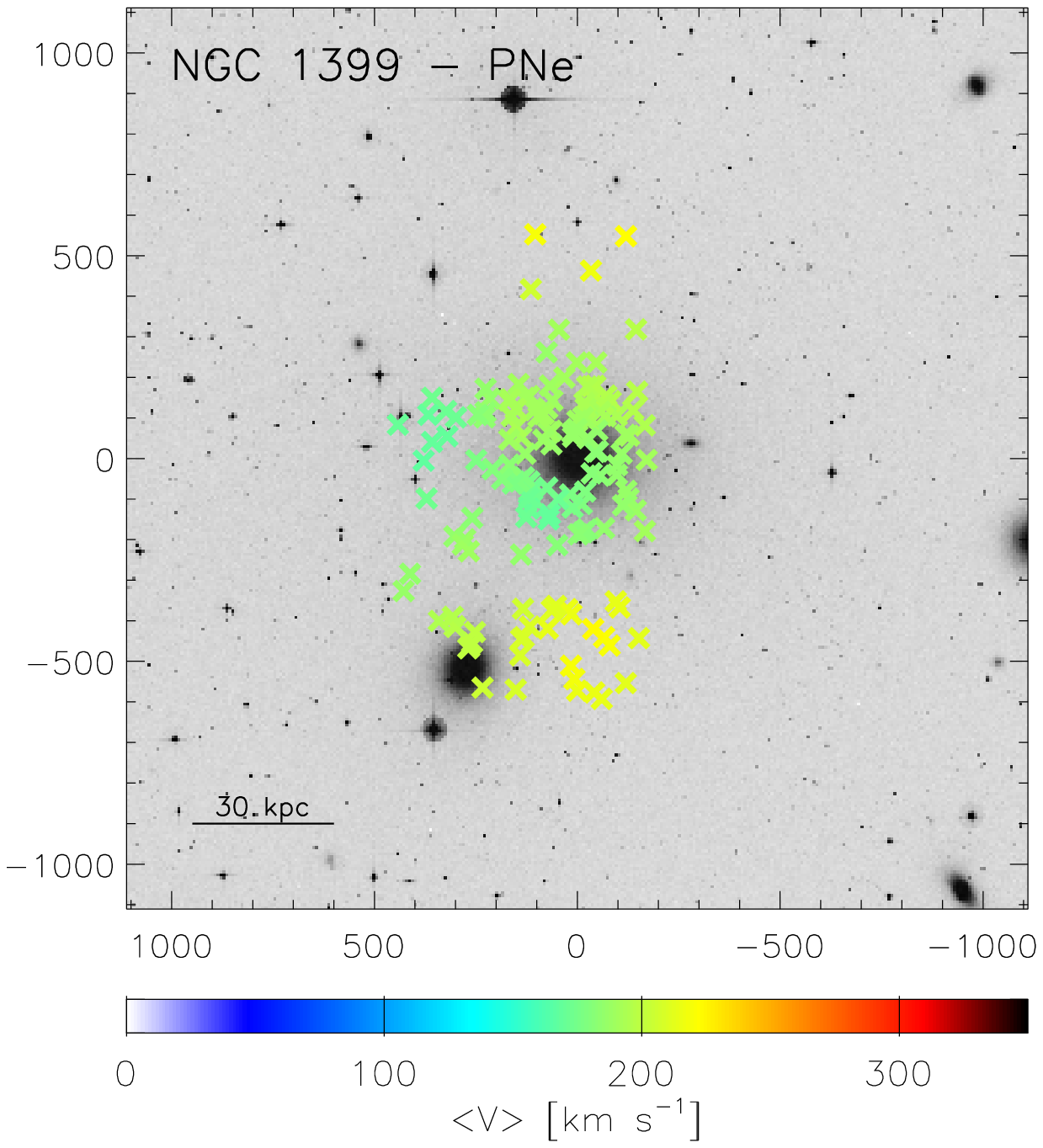,width=5.2cm,clip=,bb=53 400 404 733}
     \psfig{file=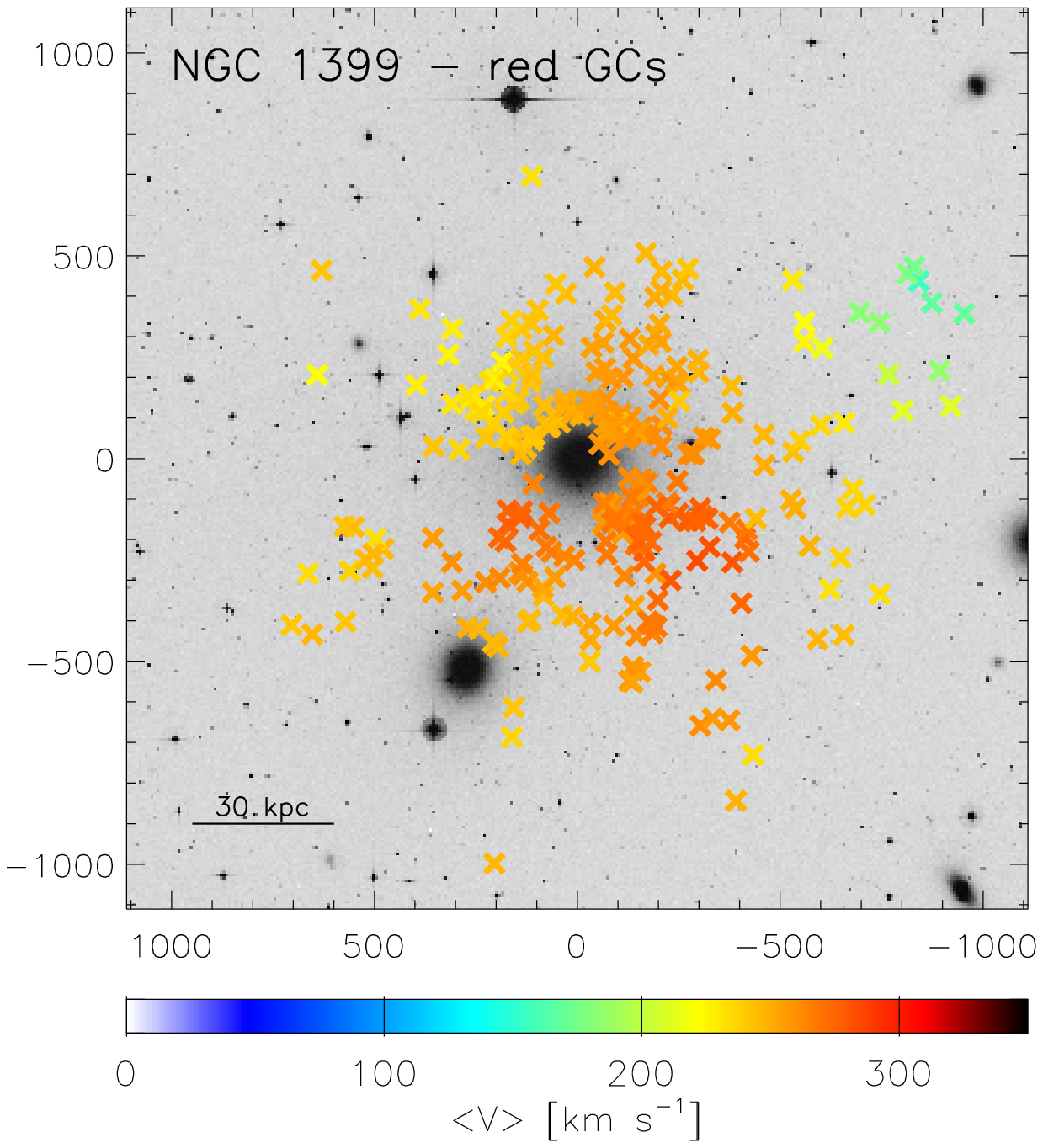,width=5.2cm,clip=,bb=53 400 404 733}
     \psfig{file=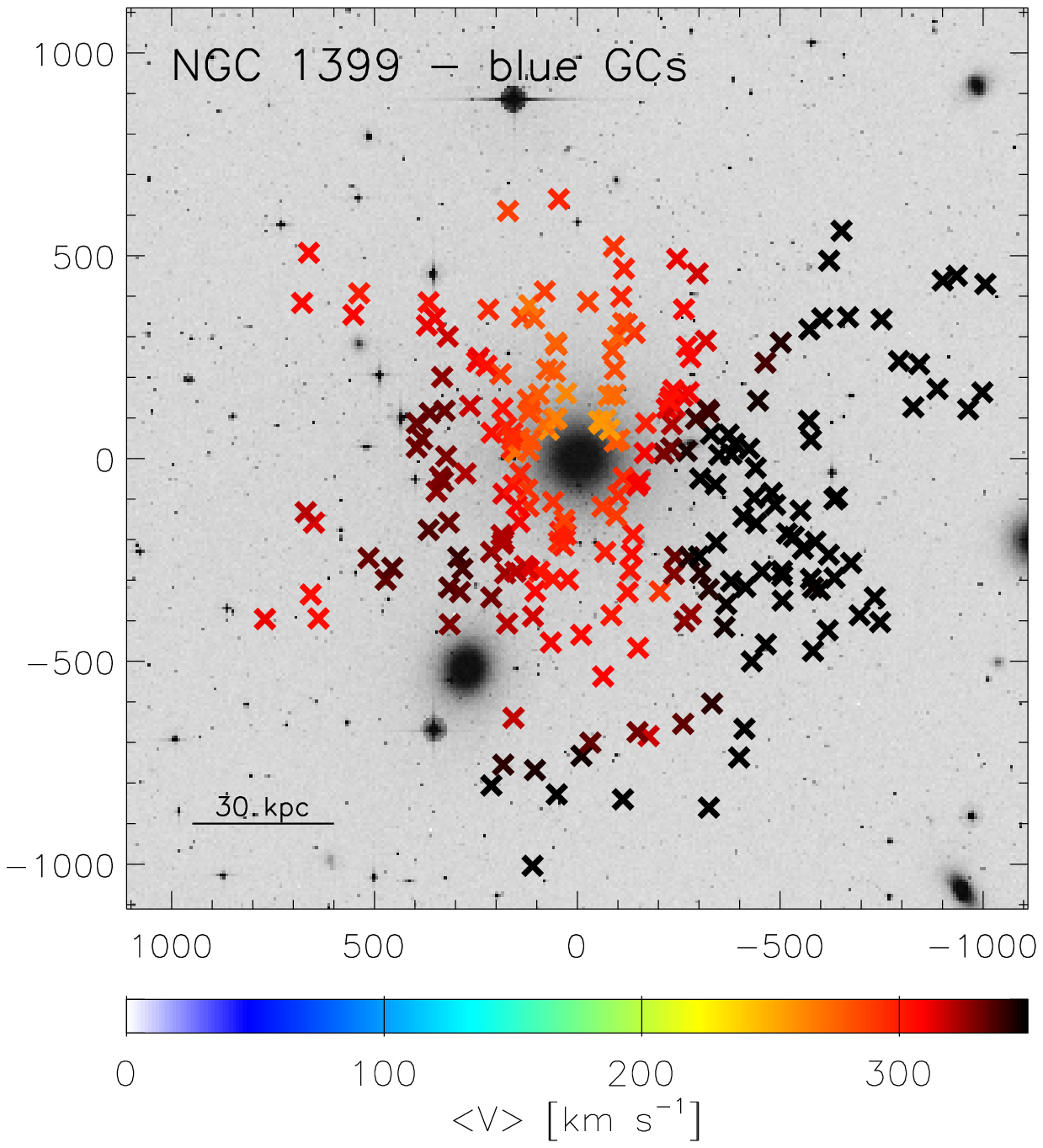,width=5.2cm,clip=,bb=53 400 404 733}
     \psfig{file=1399.pne.sigma.ps,width=4.8cm,clip=,bb=74 330 404 400,angle=90}}%4.8cm
   \hbox{
     \psfig{file=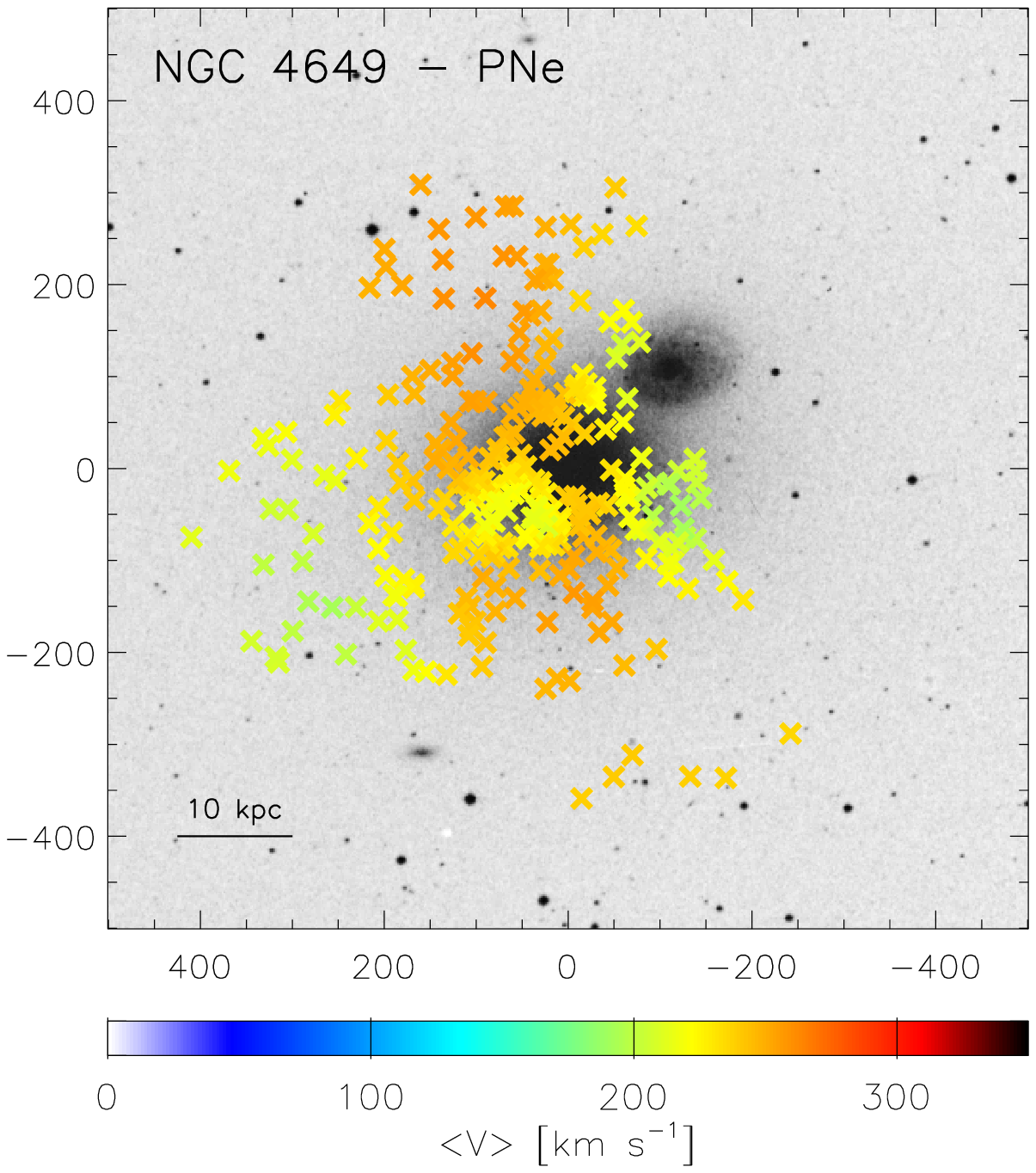,width=5.2cm,clip=,bb=53 400 404 733}
     \psfig{file=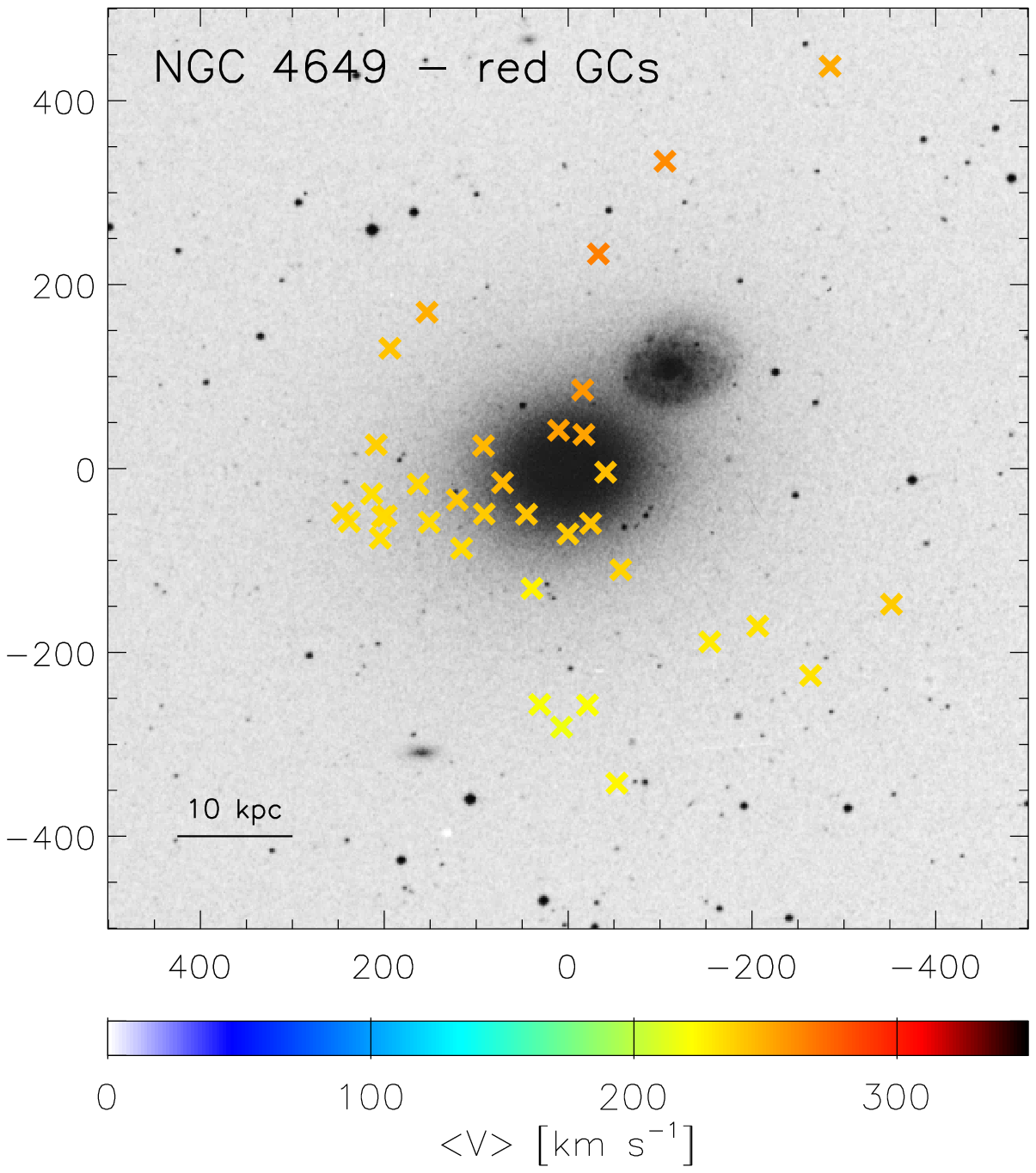,width=5.2cm,clip=,bb=53 400 404 733}
     \psfig{file=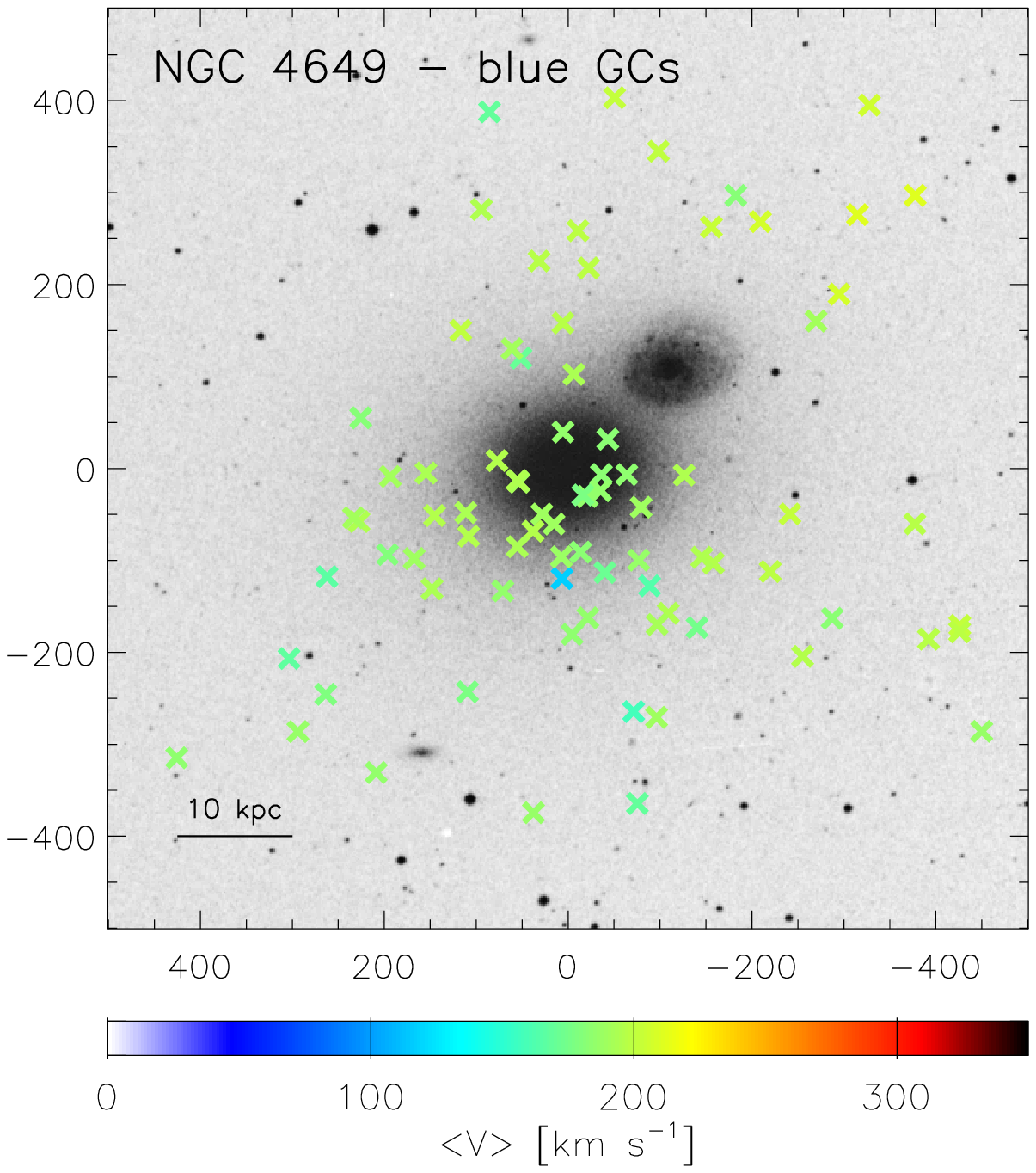,width=5.2cm,clip=,bb=53 400 404 733}
     \psfig{file=4649.pne.sigma.ps,width=4.8cm,clip=,bb=74 330 404 400,angle=90}}%4.8cm
   \hbox{
     \psfig{file=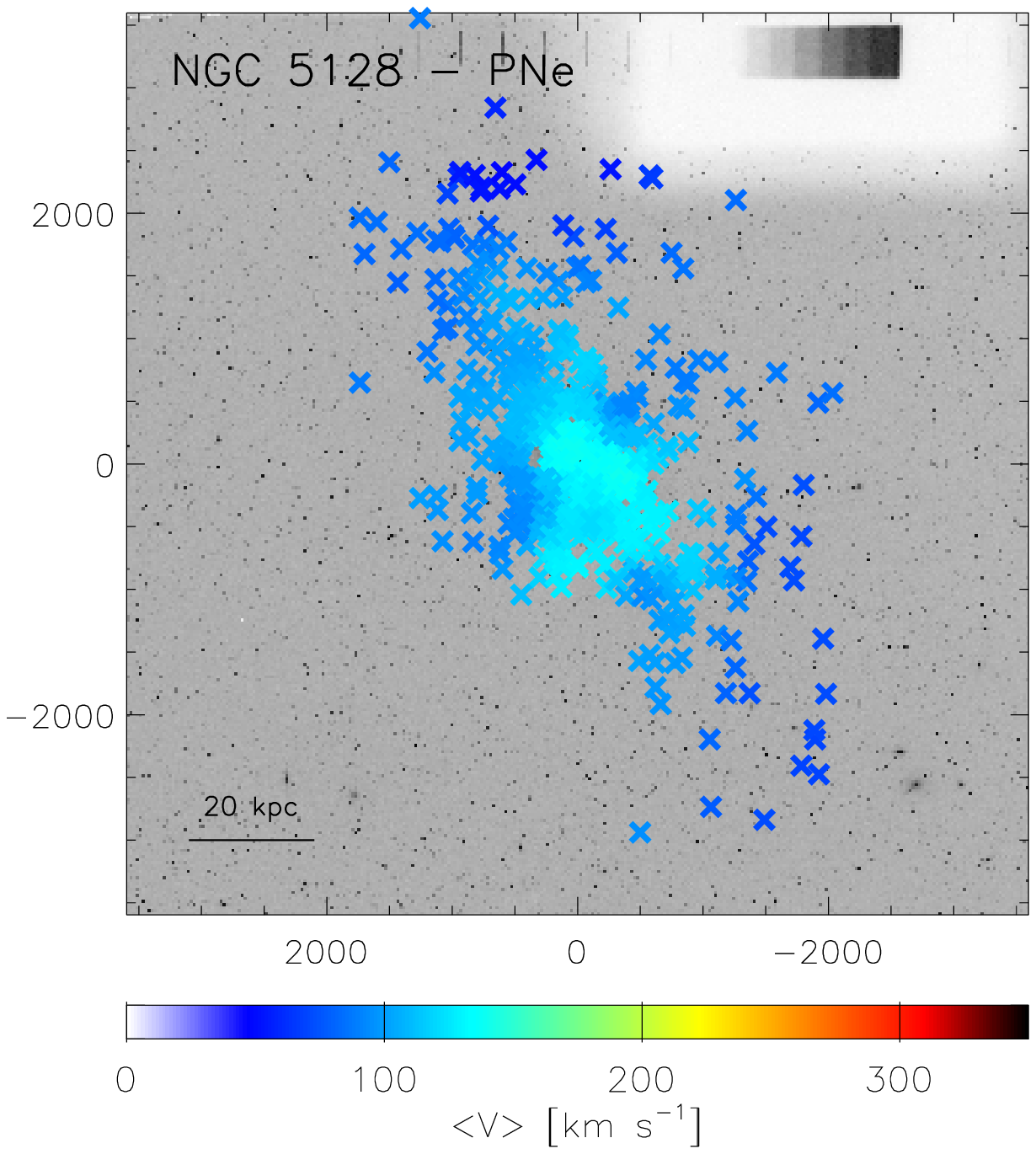,width=5.2cm,clip=,bb=53 400 404 733}
     \psfig{file=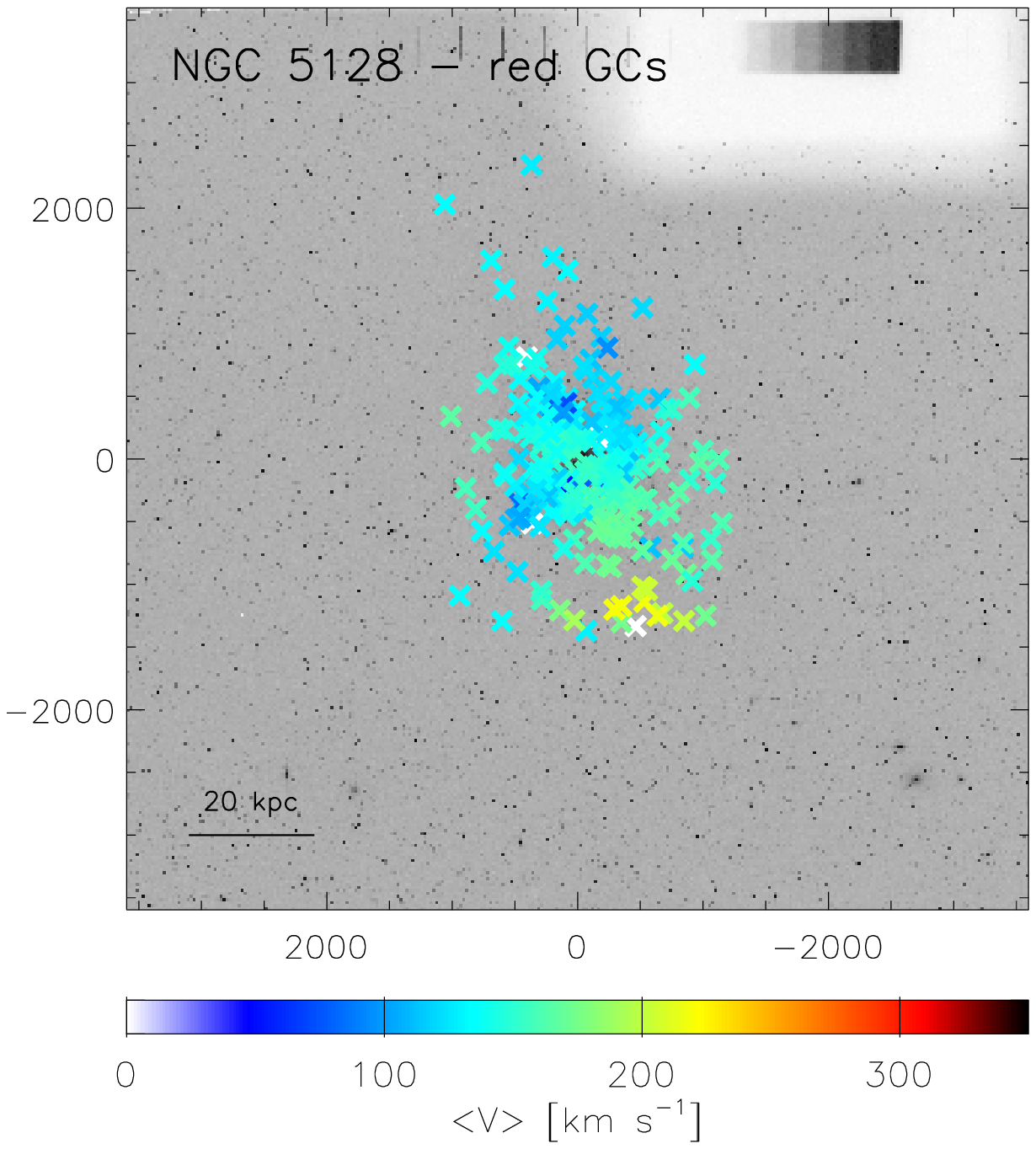,width=5.2cm,clip=,bb=53 400 404 733}
     \psfig{file=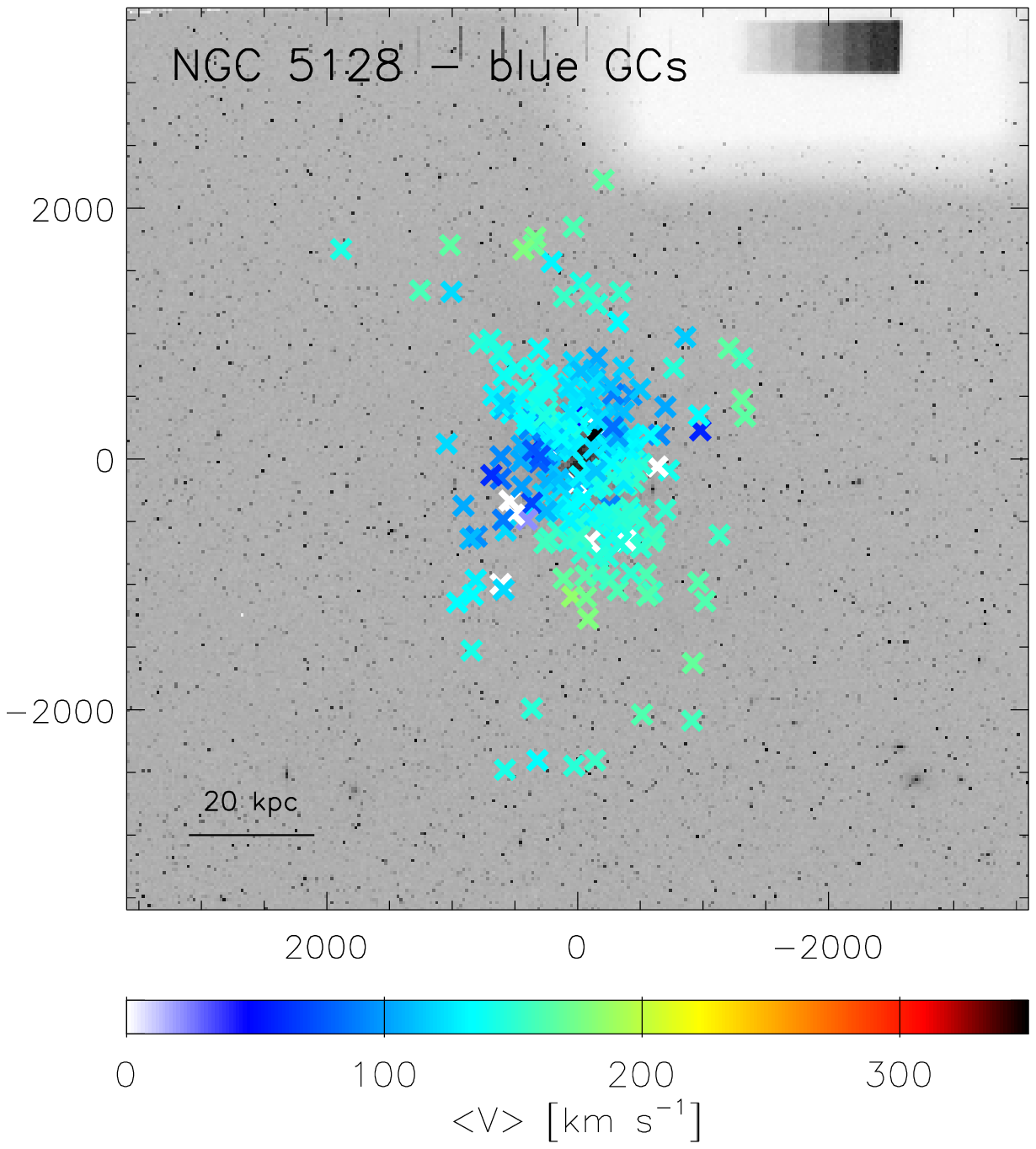,width=5.2cm,clip=,bb=53 400 404 733}
     \psfig{file=5128.pne.sigma.ps,width=4.8cm,clip=,bb=74 330 404 400,angle=90}}%4.8cm
 %    \psfig{file=1399.red.sigma.ps,width=5.6cm,clip=,bb=53 400 404 733}% 
 %    \psfig{file=4649.red.sigma.ps,width=5.6cm,clip=,bb=53 400 404 733}
 %    \psfig{file=5128.red.sigma.ps,width=5.6cm,clip=,bb=53 400 404 733}}
 %  \hbox{
 %    \psfig{file=1399.blue.sigma.ps,width=5.6cm,clip=,bb=53 330 404 733}
 %    \psfig{file=4649.blue.sigma.ps,width=5.6cm,clip=,bb=53 330 404 733}
 %    \psfig{file=5128.blue.sigma.ps,width=5.6cm,clip=,bb=53 330 404 733}}% 
 }
 \caption{Same as Fig. \ref{fig:fields} but for the mean velocity
   dispersion $\langle \sigma \rangle$.}
 \label{fig:fields_sigma}
 \end{figure*}

Recently, the distribution and kinematics of PNe and GCs in galaxy
halos have been used as first attempts to detect substructures,
  which can be interpreted as signatures of minor mergers and
accretion events that contribute to the build up of the stellar halos
in early-type galaxies \citep{Shih+10, Woodley+11, Romanowsky+12} or
to the disk heating in spirals \citep{Herrmann+09b}.

So far, a direct and detailed comparison between the two-dimensional
velocity fields inferred by GCs or PNe has been done only in few cases
(e.g. \citealt{Pota+13}). The commonly used approach was either to
combine the two datasets in the analysis, or use the most suited
dataset to study a particular science case. Little attention was paid
to the information potentially contained in the ``discrepancies''
between the PNe and GCs velocity fields, which may arise from
  differences in the progenitor population, and are worth further
  investigation.  We therefore compared the halo kinematics
independently traced by GCs and PNe in a sample of galaxies, for
  which both data are available. In this paper, we describe the
properties of three galaxies in our sample (namely NGC 1399, NGC 4649
and NGC 5128), whose two-dimensional velocity fields have kinematic
quantities that are different for these tracers. The complete
analysis of the entire sample will be presented in a future
paper. Here we concentrate on the average kinematic properties and we
do not discuss the radial distribution of individual tracers.

In this paper we will consider three aspects of the kinematics of
these tracers. In Section \ref{sec:halo} we will construct the global
velocity fields, derive the direction of rotation, and then look for
the properties of point-symmetry in the velocity fields as a test for
dynamical equilibrium. In Section \ref{sec:results} we will compare
the global kinematics of different tracers in our galaxies, and also
study specific regions of their field of view.  In Section
\ref{sec:discussion} we discuss the observed discrepancies and infer
implications on the multi-epoch mass assembly of these
galaxies. Finally, in Section \ref{sec:summary} we summarize our
results.

\section{Halo kinematics}
\label{sec:halo}

\subsection{Global kinematic maps}

We want to derive independently and compare the PNe and GCs halo
kinematics in the galaxies NGC 1399, NGC 4649, and NGC 5128. In two of
these galaxies (NGC 1399, NGC 4649) there is a satellite in the field
of view, therefore we have to account for its tracers before deriving
the global kinematic properties of the main galaxies.

\subsubsection{Separation of tracers of the satellite and galaxy}

The PNe and GCs detected in NGC 1399 and NGC 4649 contain objects
associated to the satellite galaxies NGC 1404 and NGC 4647,
respectively. Therefore, we need to separate the sample of discrete
tracers of galaxy and satellite. 
We use the procedure described in \citet{McNeil+10}, which assigns to
each tracer (PNe and GCs) a probability $P$ to belong to the
satellite, which is computed on the basis of the surface
  brightness of the galaxies at the position of the tracer and its
  velocity along the line of sight. The probability of a tracer to
  belong to the main galaxy is equal to $1-P$.

 The probability $P$ is computed as the product of two terms.  The
  first term
 is the ratio of the surface brightness of the satellite and the
 total surface brightness at the projected position of the tracer on
 the sky. This assumes that the spatial density of the tracer is
 proportional to the galaxy surface brightness
 (e.g. \citealt{Coccato+09, McNeil+10, Pota+13, Cortesi+13a}). We
 consider the proportionality factors to be the same for galaxy and
 satellite\footnote{The decomposition is robust for variations of the
   proportional factors up to 20\%.}.  The second term
 quantifies the probability that a tracer of a given velocity belongs
 to the Line of Sight Velocity Distribution (LOSVD) of the satellite.
 The LOSVDs of the galaxy and the satellite are obtained by fitting
 the observed LOSVD with two Gaussian functions (see also Figure 10 in
 \citealt{McNeil+10}). The probability $P$ is then normalized in the
 sense that each tracer is either a member of the galaxy or the
 satellite.  
Measurement errors on
 the velocity of the tracers and the surface brightness of the
 galaxies at the positions of the tracers have an impact on $P$ of the
 order of $\Delta P =5\%$, as estimated from Monte Carlo simulations.
%

%\clearpage 

\begin{figure*}
\vspace{-1cm}
  \hbox{
   \psfig{file=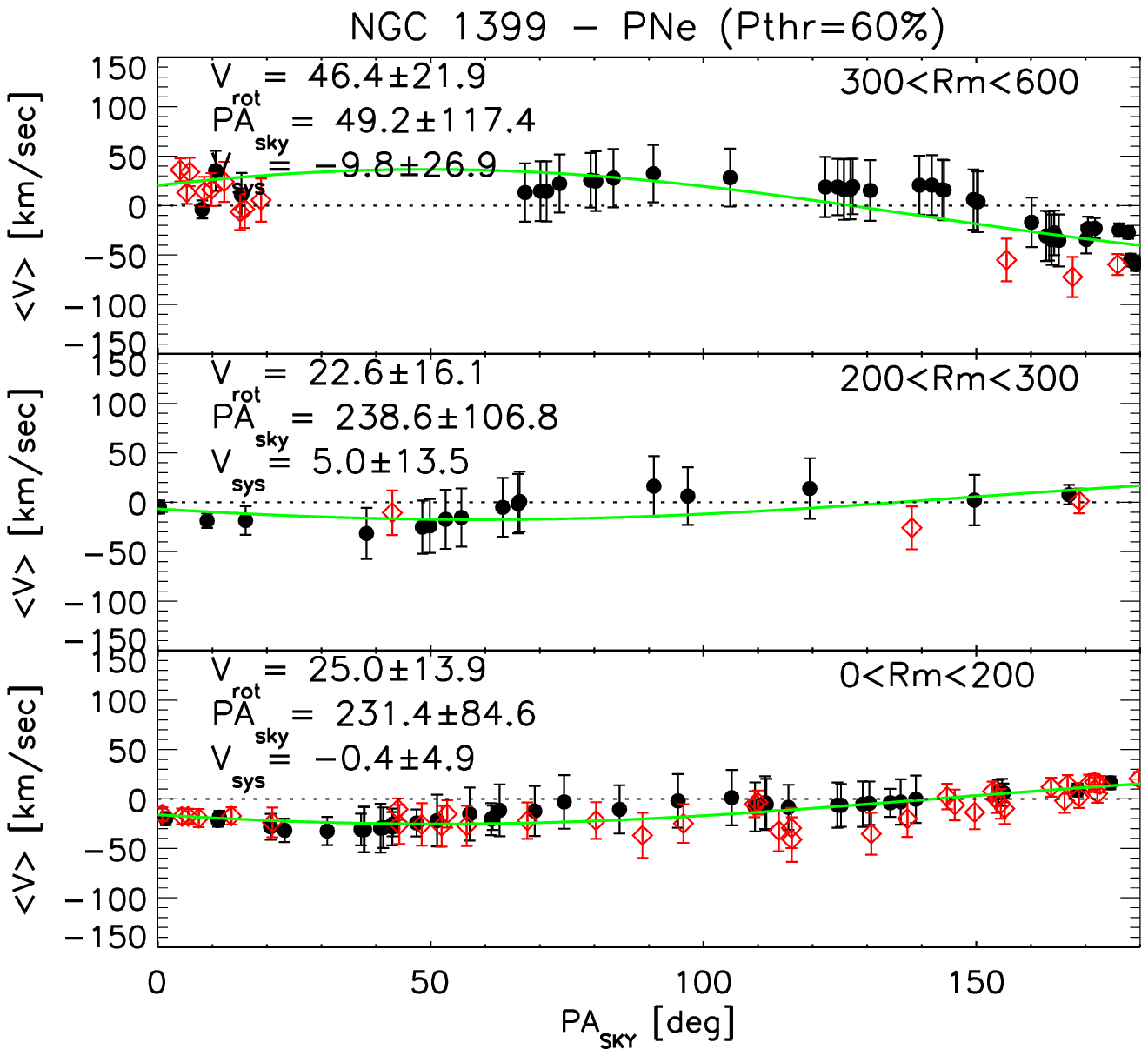,width=6cm,clip=,bb= 55 100 482 670}
    \psfig{file=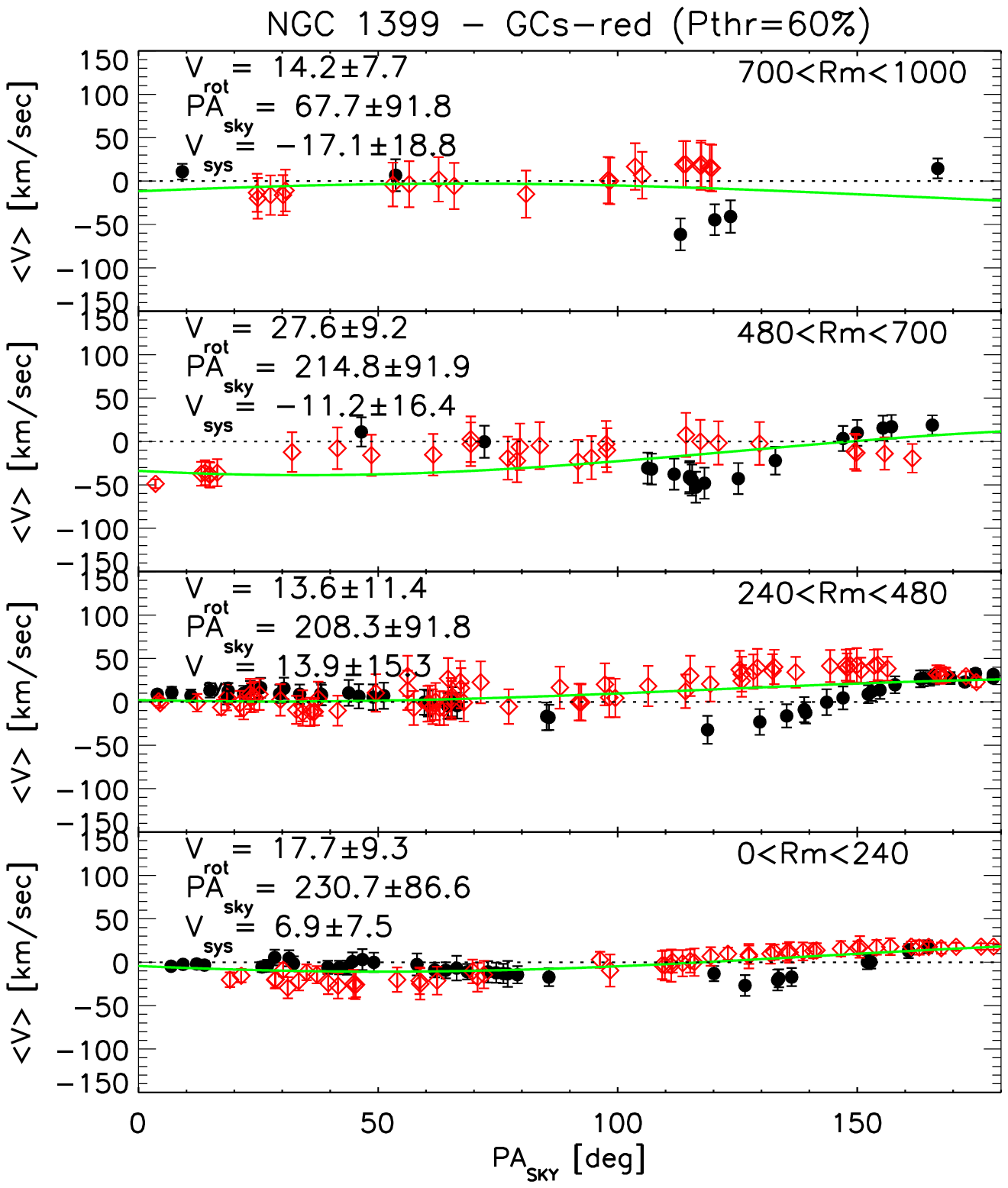,width=6cm,clip=,bb= 55 170 482 670}
    \psfig{file=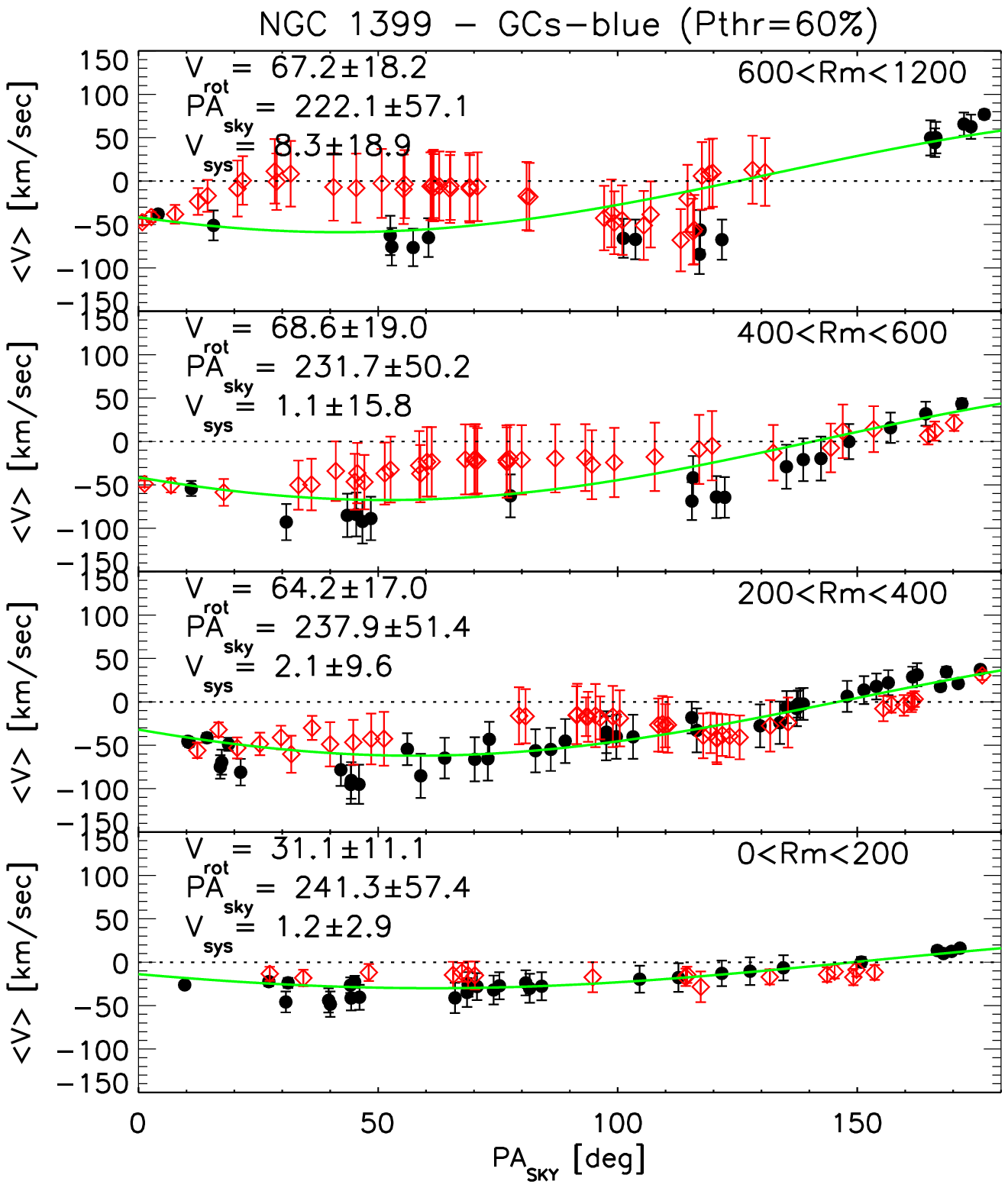,width=6cm,clip=,bb= 55 170 482 670}
    }
\caption{Mean rotation velocity of PNe (left panel), red GCs (central
  panel), and blue GCs (right panel) in several radial bins, plotted
  as a function of position angle. Angles are measured from North
  towards East. Mean velocities $\langle V \rangle$ have the
    galaxy systemic velocity subtracted. Black symbols represent
  objects at one side of the galaxy ($0^{\circ}<PA<180^{\circ}$), red
  symbols represent objects at the opposite side of the galaxy
  ($180^{\circ}<PA<360^{\circ}$), which were folded around
  $V_{sys}$. The green line is the best fit model from Equation 1. }
\label{fig:1399.rot.fold}
\end{figure*}

\begin{figure*}
\vspace{-3cm}
  \hbox{
   \psfig{file=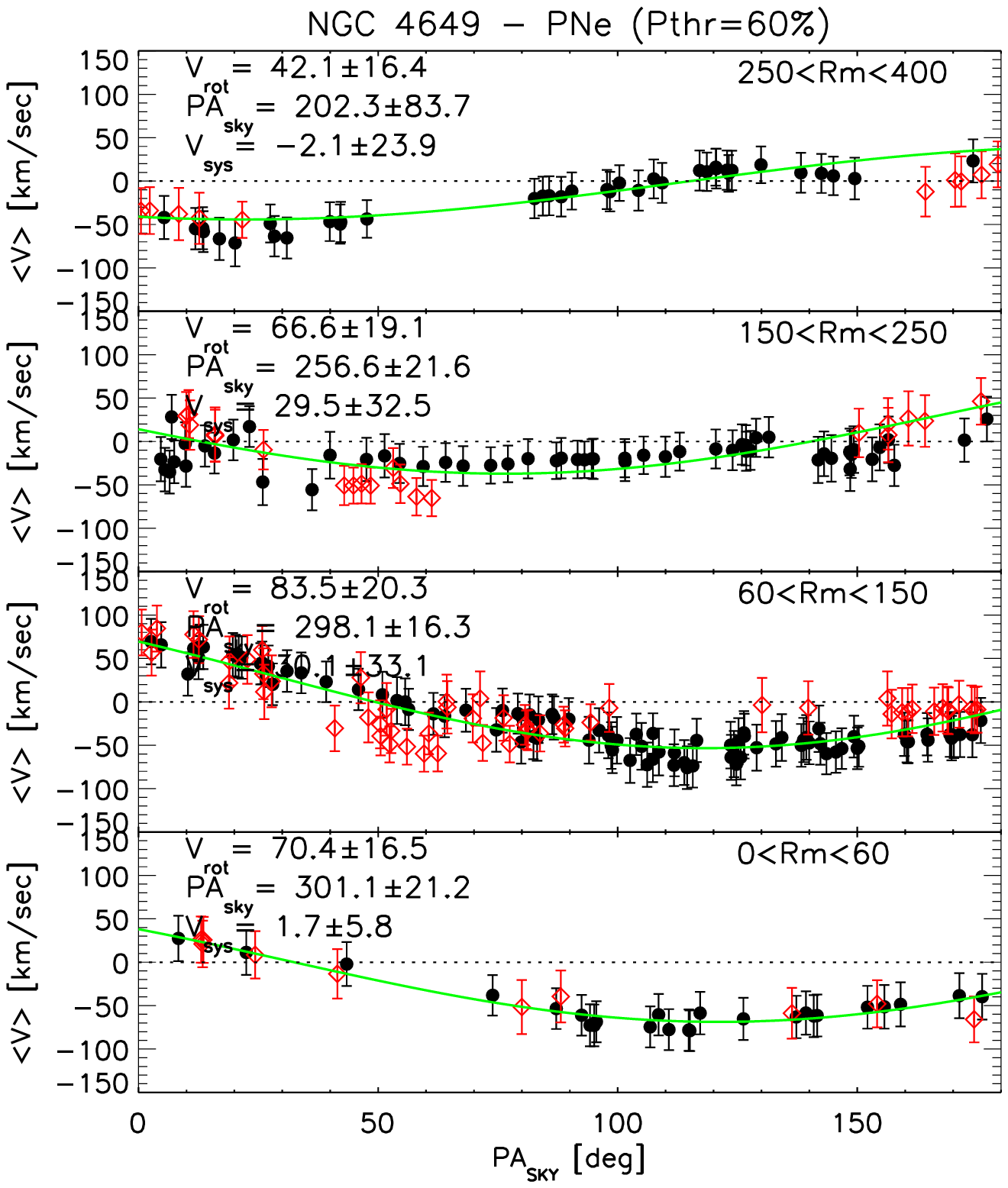,width=6cm,clip=,bb= 55 170 482 670}
    \psfig{file=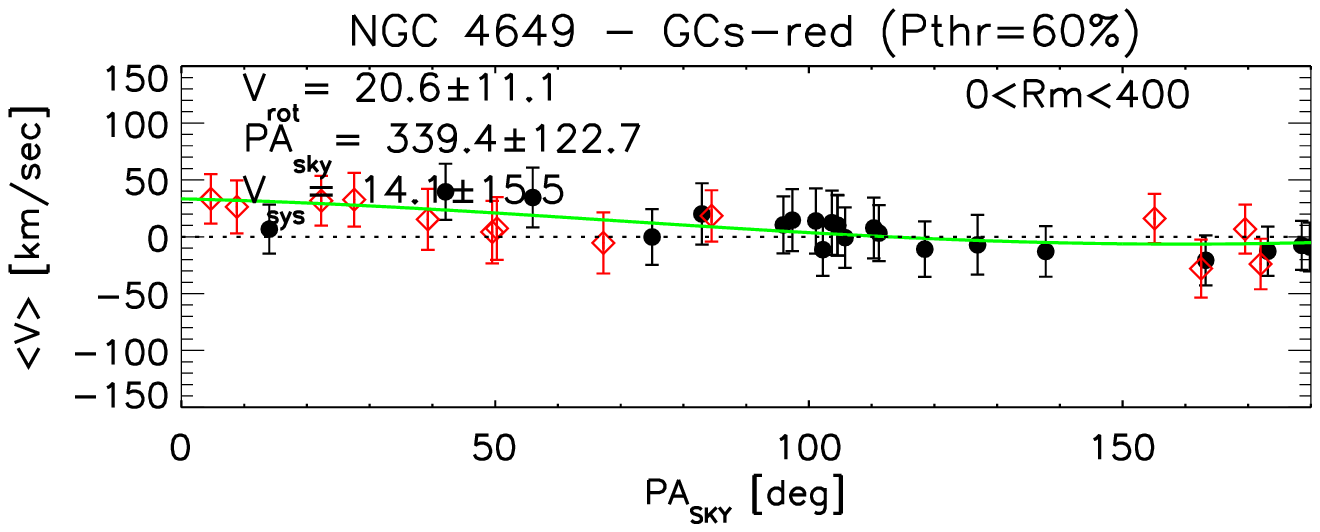,width=6cm,bb= 55 -50 482 670}
    \psfig{file=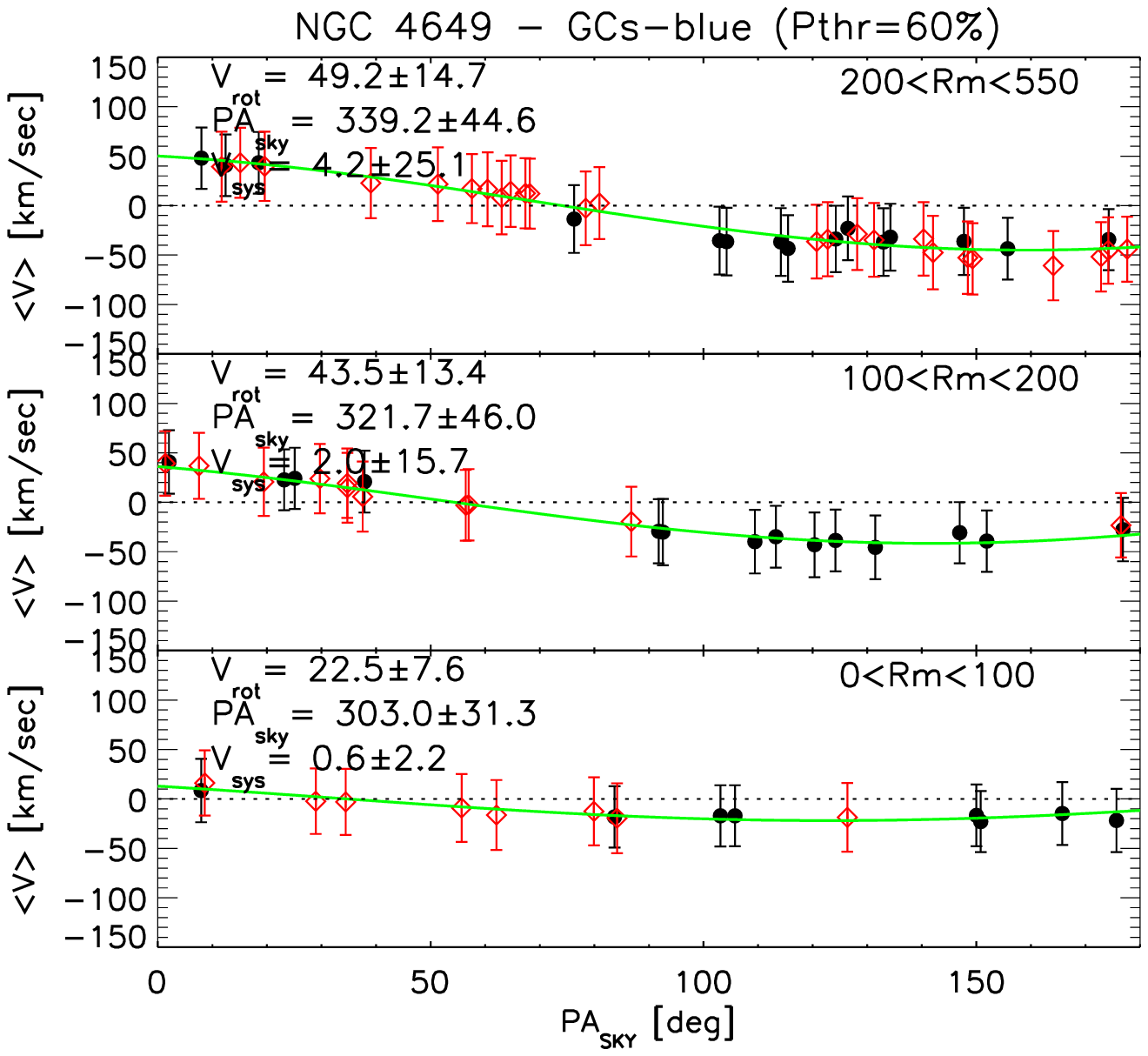,width=6cm,bb= 55 50 482 670}
    }
\caption{As in figure \ref{fig:1399.rot.fold}, but for NGC 4649.}
\label{fig:4649.rot.fold}
\end{figure*}

\begin{figure*}
\vspace{-2cm}
  \hbox{
    \psfig{file=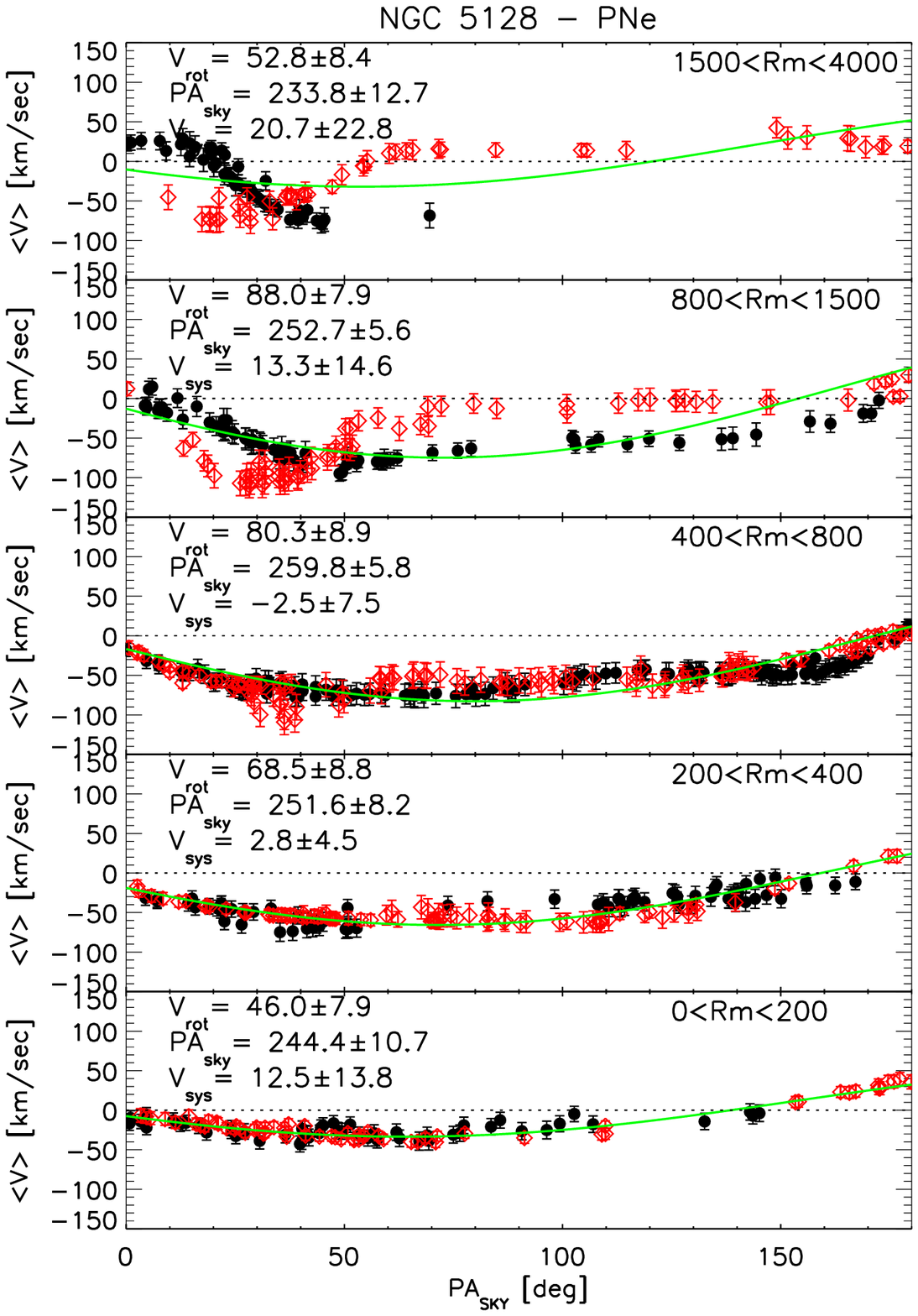,width=6cm,clip=,bb= 55 187 482 800}
    \psfig{file=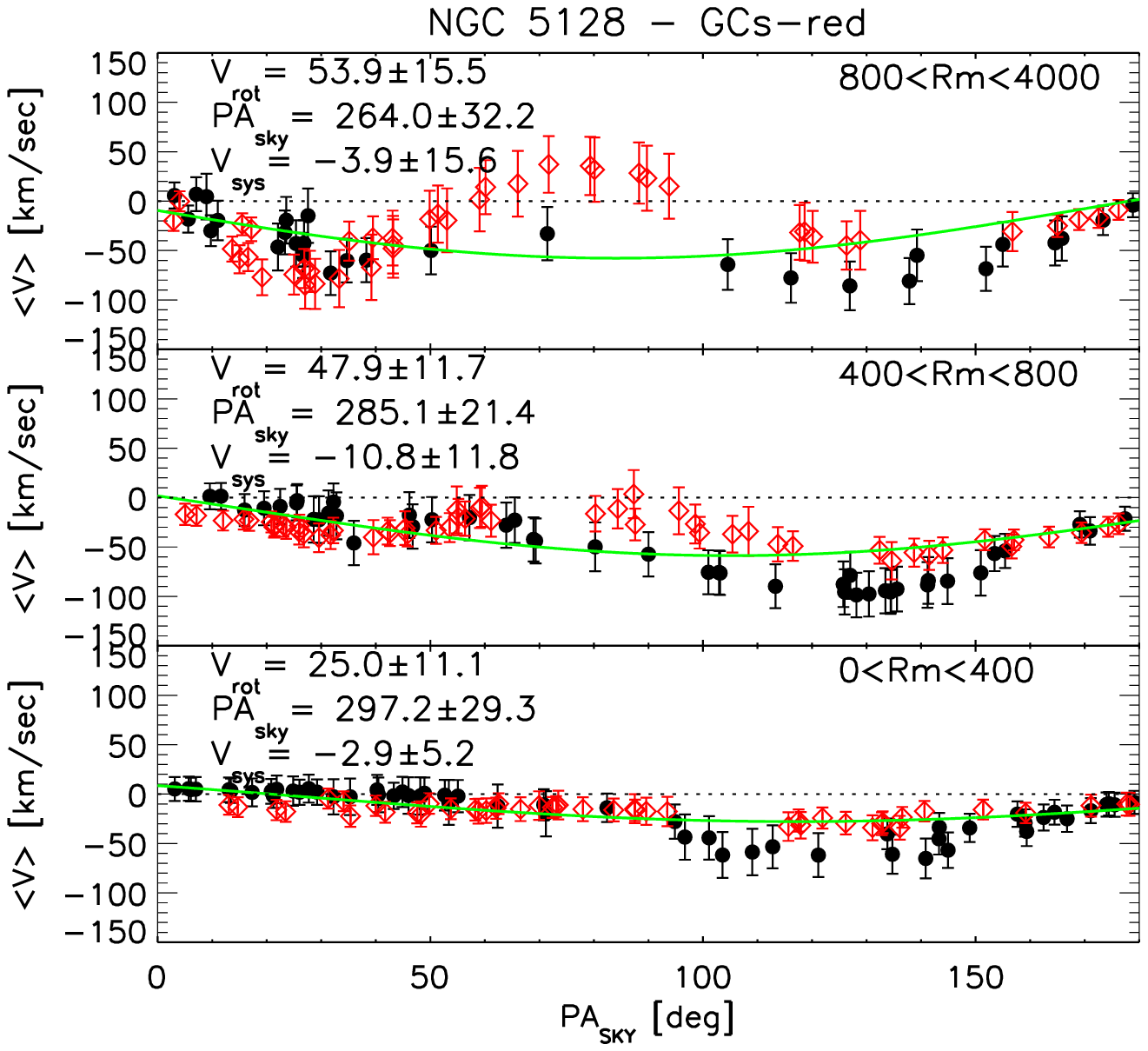,width=6cm,clip=,bb= 55 50 482 800}
    \psfig{file=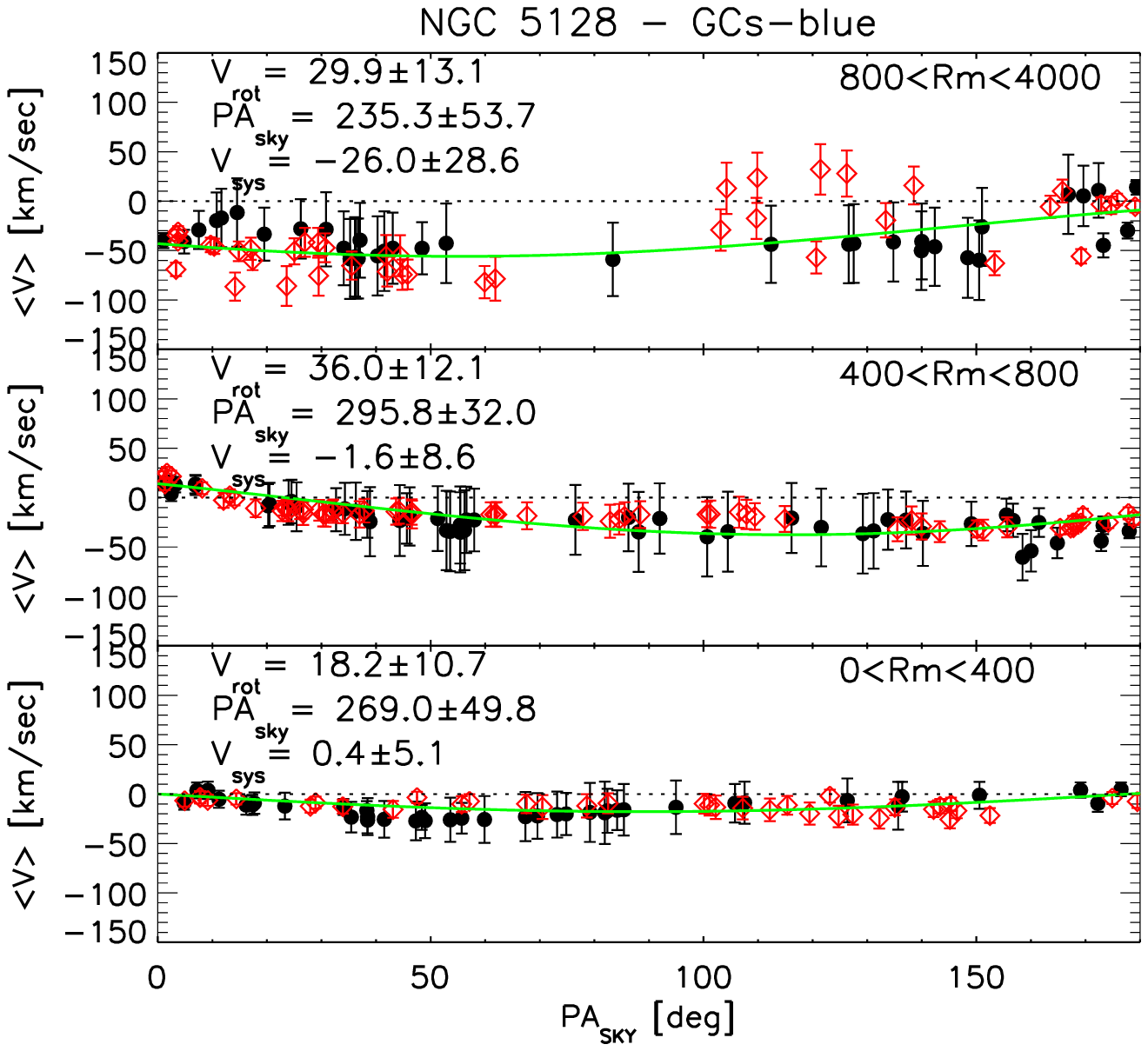,width=6cm,clip=,bb= 55 50 482 800} }
\caption{As in figure \ref{fig:1399.rot.fold}, but for NGC 5128.}
\label{fig:5128.rot.fold}
\end{figure*}

 According to this method, tracers with {\it low values} of $P$
  are likely members of the galaxy, whereas tracers with {\it high
    values} of $P$ are likely members of the satellite.

We decide to exclude from the analysis of the galaxy global kinematics
all the tracers that have probability $P> P_{\rm thr} =60\%$ (see
Figure \ref{fig:membership}). The adopted probability threshold is a
compromise that removes the tracers that are likely bound to the
satellite from the final catalog. It includes i) tracers that are
likely bound to the main galaxy, and ii) tracers that may have been
previously stripped from the satellite, have moved away from the
phase-space location of the progenitor galaxy, have not phase mixed
yet into the main galaxy potential, and thus trace a recent
interaction.  Using a higher $P_{\rm thr}$ may leave a larger number
of satellite members in the catalog, whereas a lower $P_{\rm thr}$ may
leave only tracers that are securely bound to the main galaxy, and
exclude those associated with recent accretion episodes.
Although we choose a particular value for probability threshold
($P_{\rm thr}$ = 60\%), the main results found in this work are
robust, as discussed in Appendix A.

\subsubsection{Constructing the global velocity and velocity dispersion fields}
 
 We now use the tracers associated with the galaxy to compute the
  mean halo kinematic properties of PNe and GCs. In the case of the
  GCs, we analyze red and blue GCs independently, as they have
  different spatial and kinematic properties (e.g. \citealt{Forbes+97,
    Schuberth+10, Pota+13})\footnote{We note that red and blue
      GC populations do not necessarily represent two GC populations
      with different metallicity; in fact, different features in the
    horizontal branch of a GC can led to different total colors for
    the same metallicity (e.g. \citealt{Chies-Santos+12}).}.

We reconstruct the mean two-dimensional velocity $\langle V \rangle$
and velocity dispersion $\langle \sigma \rangle$ fields for the
tracers using an adaptive Gaussian kernel smoothing technique. The
procedure is fully described and tested in \citet{Coccato+09}, and
here we provide a short description. First, we remove 3$\sigma$
velocity outliers iteratively through the {\it friendless} algorithm
as in \citet{Merrett+06}, and then we convolve the measured velocities
with a Gaussian kernel. The smoothing kernel defines the spatial
resolution, and it is chosen to get a good compromise with noise
smoothing. Because the distribution of the kinematic tracer (either
PNe or GCs) is not uniform across the field of view, we allow the
kernel to vary as a function of the local tracer's number density. In
this way, regions with low number density have a kernel larger than
that used in regions with high number density, minimizing the
  effects of spatial incompleteness of the tracers.  We allow the PNe
and GCs to have different kernels, but we adopt the same kernel for
the red and blue GC sub-populations.
We do not assume any particular symmetry of the data so as not to
  bias the analysis.

The computed mean two-dimensional velocity and velocity dispersion
fields of the PNe, the red, and blue GC sub-populations, are shown in
Figures \ref{fig:fields} and \ref{fig:fields_sigma}. The mutual
contamination between the red and blue GC sub-populations caused by
their overlapping color distributions is discussed in Appendix B.

\subsection{Rotation properties with radius}

We measure the amount and direction of rotation as a function of
radius for each tracer (PNe, red and blue GCs) in each
galaxy by fitting the following function in different radial bins:

\begin{equation}
\langle V \rangle (\phi; R)  = V_{\rm sys}(R)+V_{\rm rot}(R) \cdot \cos\left[\phi - PA_{\rm kin}(R)\right] 
\label{eqn:kinem}
\end{equation}

where $R$ is the distance of the tracer (PNe or GCs) from the galaxy center,
$\phi$ is its position angle on the sky, $\langle V \rangle$ is its
mean velocity at a given position the sky.

The fit variables for each radial bin are: $V_{\rm rot}$ (amplitude of
rotation), $V_{\rm sys}$ (mean velocity), and $PA_{\rm kin}$
(kinematic position angle). $V_{\rm sys}$ is found to be consistent
with the galaxy systemic velocity at all radii within the errors; this
suggests that the measured kinematics are not biased by spatial
incompleteness.  Errors on the best fit quantities $V_{\rm
  sys}$, $V_{\rm rot}$, and $PA_{\rm kin}$, are computed by means of
Monte Carlo simulations, as follows.
For each galaxy and each tracer, we create 500 mock catalogs with
simulated tracers at the PNe, red and blue GCs observed
  positions. In each mock catalog, the $i-$th tracer has simulated
velocity $V_{i, {\rm LOS}}^{\rm sim}$ that is drawn from a Gaussian
distribution with mean $\langle V(\phi_i; R_i) \rangle$ and standard
deviation that includes the local velocity dispersion $\langle
\sigma(\phi_i; R_i) \rangle$ and measurement errors. We then use
  the simulated catalogs to generate 500 best fit parameters, whose
standard deviations define the errors of our variables.  Therefore,
the errors computed in this way account for: i) the local
velocity dispersion of the galaxy, ii) errors in the velocity
measurements of the individual objects; and iii) the incompleteness or
inhomogeneity of the objects distribution on the sky.

\subsection{Point symmetry as a test for equilibrium}

As a further step in the analysis, we test the point-symmetry of the
computed mean velocity fields to detect deviations from dynamical
  equilibrium in our galaxies.  In dynamical equilibrium, objects on
one side of the galaxy should have velocities symmetric with respect
to objects on the opposite side ($\langle V \rangle (x,y) = -\langle
V \rangle (-x,-y)$,  with respect to the systemic velocity).

In Figures \ref{fig:1399.rot.fold}, \ref{fig:4649.rot.fold}, and
\ref{fig:5128.rot.fold} we plot the mean velocities $\langle V
\rangle$ computed at the positions of each tracer against their
position angles in each radial bin. Mean velocities are folded with
respect to the value of $V_{\rm sys}$ computed in each bin to highlight
asymmetries in the kinematics and therefore to identify those regions
where the population of tracers may deviate from dynamical
equilibrium. In these figures, we also show the best fit from Equation
\ref{eqn:kinem} for each radial bin.

\section{RESULTS}
\label{sec:results}

In the following sections, we present the results of the
  comparison of the kinematics of PNe, red and blue GCs in the three
  galaxies of our sample.

\subsection{NGC 1399}
\label{sec:1399}

 PNe data and the systemic velocity of the PN system are from
  \citet{McNeil+10}, GC data and the corresponding systemic velocity
  are from \citet{Schuberth+10}. We divide the GCs in NGC 1399 in red
  and blue populations adopting a magnitude limit of $m_R = 21.1$ and
  color separation $C-R = 1.55$ from \citet{Schuberth+10}. We now
  discuss their properties in turn.

The PN system is dominated by random motions (Figs. \ref{fig:fields},
\ref{fig:fields_sigma}). In the innermost $R<200''$, PNe are
  characterized by $V_{\rm rot} = 25 \pm 14$ \kms\ and a kinematic
  major axis $PA_{\rm kin} = 231\deg \pm 85\deg$, consistent with the
galaxy photometric major axis ($PA=290\deg$,
\citealt{Dirsch+03}). As we go to the outer regions, $R>300''$,
  PNe are characterized by $V_{\rm rot} = 46 \pm 22$ \kms\ and a
  kinematic major axis $PA_{\rm kin} = 49\deg \pm 117\deg$, nearly
  along opposite direction to that of the inner PNe (Figure
  \ref{fig:1399.rot.fold}). The observed kinematic decoupling between
  the PNe in the inner and outer regions is in agreement with the
  results by \citet{McNeil+10}, and it does not depend on the adopted
probability threshold $P_{\rm thr}$ (see Appendix A).

 The red GC population shows an asymmetric velocity field, see
  Figure \ref{fig:fields} where it is difficult to identify a clear
direction of rotation. 
The asymmetry of the velocities is more evident in Figure
\ref{fig:1399.rot.fold}, where the mean velocities $\langle V \rangle$
computed at the position of each tracer are plotted against the
  position angles: for $R>240''$ the red GCs velocity deviates from
  point-symmetry.
This suggests that the red-GC system has not reached dynamical
equilibrium yet.

The blue-GC system display less asymmetry than the red GCs
  within $400''$. Using Equation \ref{eqn:kinem}, we find rotation of
  $V_{\rm rot}=79\pm 29$ \kms\ in the radial range $4' < R \leq 8'$,
  consistent with \citet{Schuberth+10}. For larger radii ($R>400''$)
  blue GCs also deviate from point symmetry.

Rotation in the innermost $200''$ is small, $V_{\rm rot} = 31 \pm 11$
\kms, whereas it increases up to $V_{\rm rot} =67 \pm 18$ \kms in the
outskirts. The kinematic position angle remains constant ($\sim
233\deg$) with radius within the error bars, consistent with that of
the inner PN system.

\begin{figure}
%~/data/halos/kinematics/substructures_1399/substructures.pro  
%\hspace{-.3cm}
\begin{center}
\vbox{
  \psfig{file=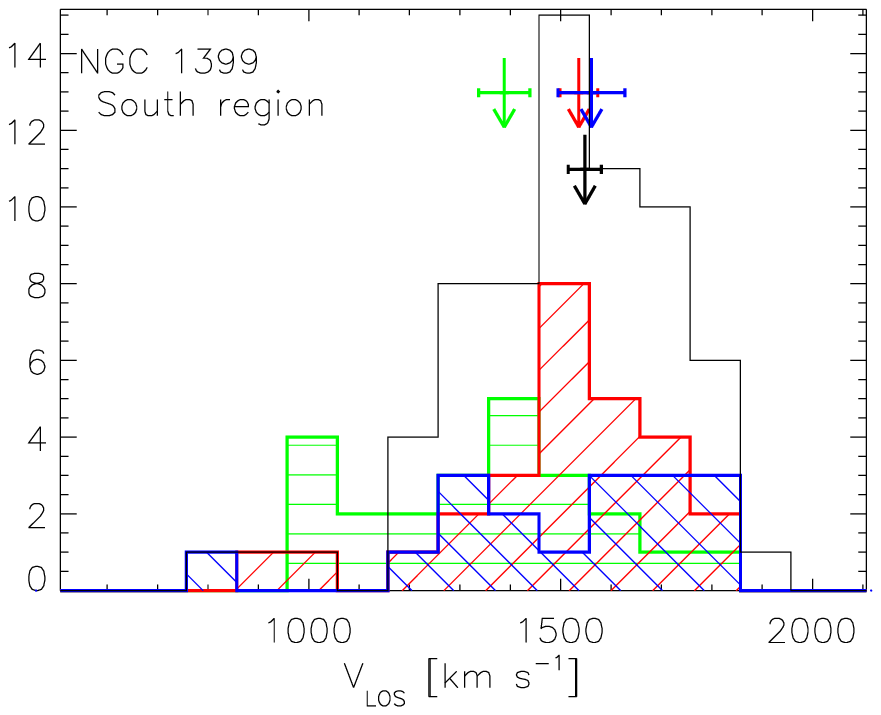,width=6.0cm,clip=}
  \psfig{file=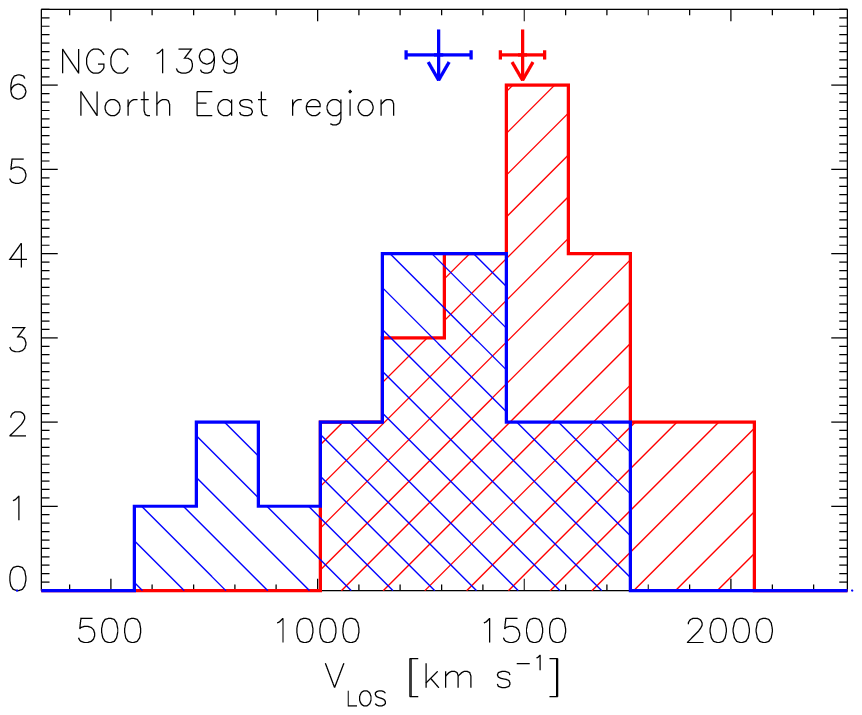,width=6.0cm,clip=}
  \psfig{file=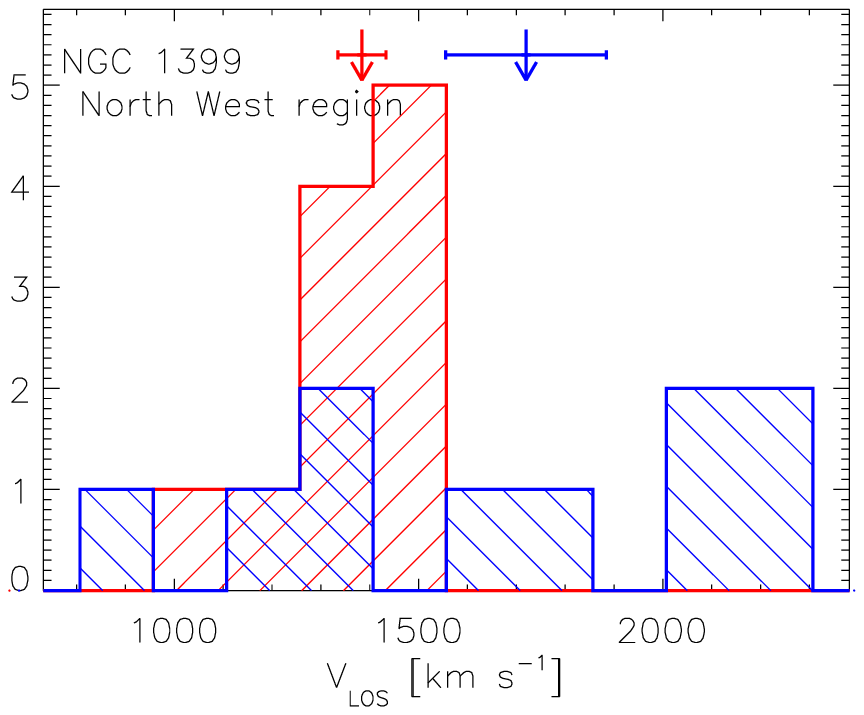,width=6.0cm,clip=}}
\caption{Velocity distribution of PNe (green), GCs (black), red-GCs
  (red) and blue-GCs (blue) in the South (upper panel), North East
  (central panel), and North West regions of NGC 1399. Arrows mark the
  median radial velocities of the systems with the same
  colors. Velocities are in the heliocentric reference system.
    The color is not defined for all the GCs, therefore the total
    number of GCs is larger that the combined blue and red GC
    populations.}
\label{fig:hist}
\end{center}
\end{figure}

\subsubsection{Local velocity differences in the field of view}

 There are also ``local'' differences among PNe and GCs kinematics
  in NGC 1399, in addition to those among the mean velocity fields
  $\langle V \rangle$.  Indeed, there are regions in the field of
view where the difference between the kinematics of the different
tracers is more pronounced. They are located at North-East, North-West
and South of NGC 1399, see locations in Fig. \ref{fig:fields}. We
show in Figure \ref{fig:hist} the line-of-sight velocity distributions
of PNe and GCs in these regions, and quantify their differences below.

\begin{itemize}

\item In the S-region, the PNe and GCs have median velocities that
  differ at 81\% confidence level ($\simeq 1.3 \sigma$). From a
  Kolmogorov-Smirnov test \citep{Press+02}, the probability that PNe
  and GCs are drawn from the same distribution is $1 \pm 1$\%. 
    The error on the probability is computed by a set of MC simulations
    that account for errors in $V_{\rm LOS}$.  The probability depends
  marginally on the GCs color. 

The probability that PNe and red GCs are drawn from the same
distribution is $1 \pm 1$\%, this value is robust towards
  contamination from the blue GC population. The probability that
PNe and blue GCs are drawn from the same distribution is $8 \pm 5$\%,
 and depends on the amount of contamination from the red GCs (see
  Appendix B).  The red and blue GCs have higher probability to be
drawn from the same distribution ($69 \pm 22$\%).

\item In the NE-region, the probability that red and blue GCs are
  drawn from the same distribution is $3 \pm 2$\%. Their median
  velocities differ at 77\% confidence level ($\simeq 1.2 \sigma$) ,
  and they have consistent velocity dispersion. No PNe are detected in
  this region.

\item In the NW-region, the probability that red and blue GCs are
  drawn from the same distribution is $3 \pm 4$\%. Their median radial
  velocities differ at 77\% confidence level, and their velocity
  dispersion differ at 98\% confidence level ($\sim 2.3 \sigma$). The
  standard deviation of red GCs in the NW region is $\sigma_{\rm RED,
    NW}=153 \pm 33$ \kms, whereas the standard deviations of the blue
  GCs in the NW region is $\sigma_{\rm RED, NW}=471 \pm 105$ \kms. No
  PNe are detected in this region.

  GCs in the NW region might also be contaminated or perturbed
  by the presence of NGC 1387, which is 20 arcmin to the west of
    NGC 1399 and at a systemic velocity of 1328 \kms, which is very
  close to the systemic velocity of NGC 1399 (1447 \kms).
  The fact that blue GCs have higher velocity dispersion than the red
  GCs in this region indicates that the blue GC system could be
    more contaminated by the presence of objects from NGC 1387.
  Unfortunately, the photometric data of NGC 1399 and NGC 1387 are not
  extended enough to allow a membership determination like the one done for
  NGC 1404.

\end{itemize}

\smallskip

Results shown in Figure \ref{fig:hist} plot the observed $V_{\rm
    LOS}$, and are independent from the kernel smoothing technique
  used to recover the mean velocity fields. In addition, they weakly
depend on the adopted threshold probability $P_{\rm thr}$ for
  satellite membership (see Appendix A).

In summary, the kinematics of PNe, red and blue GCs in NGC 1399,
  are dominated by velocity dispersion, with global and local
  differences among their velocity fields. Furthermore, the blue and
  red GCs velocity fields deviate from point-symmetry at large radii,
  indicating that they have not reached dynamical equilibrium yet.

\subsection{NGC 4649}
\label{sec:4649}

PN data and the corresponding systemic velocity are from
\citet{Teodorescu+11}, GC data and the corresponding systemic
velocity are from \citet{Lee+08}. We divide the GC system of NGC
  4649 in red and blue GCs adopting a color separation of $C-T_1=1.65$
  mag as in \citet{Lee+08}. We now discuss their properties in turn.

 The PN system in NGC 4649 is dominated by velocity dispersion
  and it is a multi-spin system with two kinematic components at
nearly orthogonal directions of rotation (see Figs.  \ref{fig:fields}
and \ref{fig:fields_sigma}). As clearly shown in Figure
  \ref{fig:4649.rot.fold}, PNe in the inner $R<60''$ rotate with an
amplitude $V_{\rm rot}= 70 \pm 17$ \kms\ along $PA_{\rm kin} =
301\deg\pm 21\deg$, consistent with the photometric major axis
($PA=285\deg$, \citealt{RC3}); wherease PNe in the outer $R>250''$
rotate along $PA_{\rm kin} = 202\deg \pm 84\deg$ with similar
amplitude ($V_{\rm rot}= 42 \pm 16$ \kms).  The PNe velocity field
 is consistent with point-symmetry (Figure
\ref{fig:4649.rot.fold}). The two-dimensional velocity field is
similar to that shown by \citet{Das+11}, obtained using different
criteria for the removal of satellite members.

 We measure $V_{\rm rot} = 21 \pm 11$ \kms\ and $PA_{\rm
    kin}=339\deg \pm 123\deg$ for the red GC system, independently
  from the adopted $P_{\rm thr}$, and the system is pressure supported
  ($\langle \sigma \rangle \sim 250$ \kms).

The blue GC sub-population shows a regular velocity field. In the
inner $R<60''$ the blue GCs rotate along $PA_{\rm kin}=303\deg \pm
31\deg$ with an amplitude of $V_{\rm rot} = 23 \pm 8$ \kms, slower
than the PNe in the same radial range but along similar direction.
The $PA_{\rm kin}$ and $V_{\rm rot}$ change gradually with the
radius. At $R>200''$ we measure rotation $V_{\rm rot} = 49 \pm 15$
\kms\ along $PA_{\rm kin}=339\deg \pm 45 \deg$, nearly opposite to
that of the outermost PNe; see Figure \ref{fig:4649.rot.fold}.

\begin{figure}
%~/data2/halos2/kinematics2/revised2/05/simulations/do_simulations.pro   esegui;
  \psfig{file=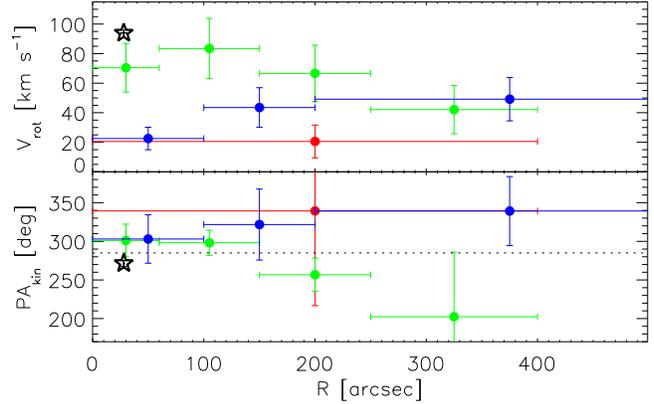,width=8.5cm,clip=,bb=33 141 366 351}
  \caption{Rotation amplitude ($V_{\rm rot}$) and direction
    ($PA_{\rm kin}$) of the stars (black star), PNe (full green
    circles), blue GCs (full blue circles), and red (full red circles)
    GC systems in NGC 4649, as obtained from Equation
    \ref{eqn:kinem}. The dotted line indicates the position angle of
    the photometric major axis. The horizontal bars represent the
    extension of the radial bins.}
\label{fig:kinem4649}
\end{figure}

In Figure \ref{fig:kinem4649} we show the value of $PA_{\rm kin}$
  and $V_{\rm rot}$ for stars, PNe, red and blue GCs in different
  radial bins. The difference in $PA_{\rm kin}$ between PNe and blue
  GCs is $\sim 150^{\circ}$ for the outer radial bin.
The orthogonal directions of rotation between outermost PNe and blue
GCs is independent from the adopted probability threshold $P_{\rm
  thr}$, whereas the amplitude of rotation mildly depends on it
(Appendix A).  The derived kinematic properties of blue GCs are not
affected by contamination from the red-GC sub-population (Appendix
B).

 In summary, the kinematics of PNe, red and blue GCs in NGC 4649,
  are dominated by velocity dispersion; with differences in the
  angular momentum vectors. PNe are a multy-spin system, with inner
  and outer regions having nearly orthogonal angular momenta.
  Furthermore, the PNe, blue and red GCs velocity fields are
  point-symmetric, and consistent with dynamical equilibrium in the
  studied radial range.

\subsection{NGC 5128}
\label{sec:5128}

 Differences of the PNe and GCs kinematics in NGC 5128 were
  previously noted by \citet{Peng+04b} and \citet{Woodley+07,
    Woodley+10b}. 
The latter found that the GC system rotates slowly, with no
significant differences between red and blue GCs. The PNs system
rotates faster along the same direction than the GC system, but they
have lower velocity dispersion in the outer regions.

Here, we re-analyze these data sets with the same method as the one
adopted for the other galaxies in our sample.  PNe data and the
  corresponding systemic velocity are from \citet{Peng+04}, GC data
and the corresponding systemic velocity are from
\citet{Woodley+10a, Woodley+10b}.  Following \citet{Woodley+10b}, we
use the GCs metallicity [Fe/H] $=-1$ as threshold to separate red
(metal rich) and blue (metal poor) GCs.

From Eq.  \ref{eqn:kinem}, we find that PNe reach a maximum rotation
$V_{\rm rot}= 88 \pm 8$ \kms\ around $20'$, and then a decline to
$V_{\rm rot}= 53 \pm 8$ \kms\ in the halo, consistent with
\citealt{Woodley+07}. The velocity dispersion is $\simeq 140$
\kms\ in the center and then it declines with radius down to $\simeq
80$ \kms.
 Within $800''$, the PNe velocity field shows disk-like rotation,
  with a variation of the kinematic position angle in the central 5
kpc due to the triaxiality of the system (see also \citealt{Peng+04}).

At radii $>800''$ the PN system is not point-symmetric: if we
  apply Equation \ref{eqn:kinem} independently to the PN sub-samples
  for the approaching and receding side of NGC 5128, the two kinematic
  position angles differ by $\sim 15^{\circ}$ (see also Figure
  \ref{fig:5128.rot.fold}). This asymmetry is clearly visible in
  Figure \ref{fig:5128.rot.fold} for the two outermost radial
  bins. This result suggests that the PN system of NGC 5128 has not
  yet reached dynamical equilibrium in the halo at these radii.

We check that this result is not an artifact due to spatial
incompleteness of the PN sample in the outer regions, as follows.
We select the PNe on the receding (approaching) side of the galaxy,
and symmetrize them onto the approaching (receding) side. We compute
the smoothed velocity $\langle V_{\rm S} \rangle$ and velocity
dispersion $\langle \sigma_{\rm S} \rangle$ fields from the
approaching and receding symmetric PN samples. We then extracted a
simulated PN data set with positions $(x,y)$ that matches the
observations, and velocities $V(x,y)$ randomly selected from a
Gaussian distribution with mean velocity $\langle V_{\rm S}(x,y)
\rangle$ and velocity dispersion $\sqrt{\langle \sigma_{\rm
    S}(x,y)\rangle^2 + dV(x,y)^2} $. These simulated data sets
  therefore have the same spatial incompleteness as the original
  observed sample. We then compute the smoothed velocity fields and
  study their point symmetry properties. The smoothed velocity fields
reconstructed from these simulated PN data turn out to be point
symmetric. We repeat the test for 100 simulated datasets, and the
  results are consistent with point symmetry. We conclude that the
  observed deviation from point-symmetry is an intrinsic property of
  the PN population in NGC 5128 and it is not caused by spatial
  incompleteness.

 Within $800''$ the velocity fields of red and blue GCs are well
  reproduced by Equation \ref{eqn:kinem}, they both show weak average rotation of
  $V_{\rm rot}= 36 \pm 8$, $V_{\rm rot}= 26 \pm 8$, respectively,
  along similar position angles ($PA_{\rm kin} = 289^{\circ} \pm
  15^{\circ}$).

The central velocity dispersion for GCs is consistent with that of PNe
($\approx 150$ \kms), but it does not decrease with radius. 
  Similarly to PNe, red and blue GCs deviates from point symmetry at
  radii $>800''$, signaling that they have not reached dynamical
  equilibrium yet. This is effect is stronger for the red GC system
  (see Figure \ref{fig:5128.rot.fold}).

In Summary, within $800''$ the current result supports a picture
  where the GCs are more related to a spheroidal component, dominated
  by velocity dispersion, while the PNe are more rotationally
  supported, and related to a disk-like component. At large radii,
  deviations from point-symmetry are observed for both PNe and GC
  systems, which indicate that the halo has not reached equilibrium
  yet.

\section{Discussion}
\label{sec:discussion}

We show in Sect. \ref{sec:results} that the PN and red and blue GC
systems in NGC 1399, NGC 4649, and NGC 5128 have different kinematic
properties.  Moreover, in the outer regions of NGC 1399 and NGC
  5128 we detect deviations from point-symmetry in their velocity
fields (Figures \ref{fig:fields}, \ref{fig:1399.rot.fold},
\ref{fig:4649.rot.fold}, and \ref{fig:5128.rot.fold}), which are not
consistent with dynamical equilibrium. The observed different
kinematic properties, including deviations from point-symmetry,
between PNe and GCs suggest that these systems have been
accreted at different times by the host galaxy.

The observed asymmetries in the velocity fields in the outer
  radial bins are consistent with an interaction/accretion event which
  occured in the last few Gyr, because the dynamical timescale in the
  halos of massive galaxies is of the order of 1 Gyr, and therefore
  dynamical equilibrium is expected to be reached in a few Gyr.
 The statement that NGC 1399 and NGC 4649 could have experienced a
  recent interaction, is supported by the fact that they are located
in dense environments and have a luminous companion nearby (NGC 1404
and NGC 4647, respectively).  Moreover, multi-epoch accretion
  events for the mass assembly of their halos are also consistent with
  the presence of the multi-spin components observed in their PN system.
 In the case of NGC 5128, this galaxy is considered a merger
  remnant (e.g. \citealt{Peng+04b, Woodley+10a, Crnojevic+13} and
  references therein) and deviations from symmetry in the outer
  regions indicate that this event took place in the last few Gyr.

 We now discuss how the observed differences between PNe and GCs
  kinematics may arise from a multi-epoch mass assembly scenario.

\begin{itemize}

\item {\it Tidal stripping.} NGC 1399 and NGC 4649 could have stripped
  material from their less massive satellites. The GC systems in
  galaxies are generally more extended than the underlying stellar
  population (e.g. \citealt{Harris91});  this implies that in some
    cases the outer GCs may be stripped in the interaction, more
    frequently than stars and PNe, which are tightly bound. It follows
    that the signature of the tidal stripping will be most visible in
    the GC system, which then traces different kinematics than PNe and
    stars.  Although our sample is limited, the above prediction is
    supported by the fact that the PNe kinematics are on average more
    point symmetric than the GC system thus suggesting that the PN system
    is less perturbed.

 \citet{Forbes+97} explored this scenario for NGC 1399. This
galaxy has an high specific frequency (i.e. number of GCs per unit
galaxy luminosity, normalized at $M_V=15$, \citealt{Harris+81}), and
GCs are preferentially distributed in the outer galaxy regions. 
Its satellite, NGC 1404, has an observed luminosity which is
  consistent with its stellar velocity dispersion. This suggests that
  only a small fraction of its stars were removed in the stripping
  process, whereas its low specific frequency for GCs indicates that
  they many of them were lost to NGC 1399 during the interaction.

Information on the GCs specific frequency is not available for NGC
4647, the companion of NGC 4649, and we therefore do not have elements
to study the tidal stripping between NGC 4649 and NGC 4647 as in the
case of NGC 1399. Nevertheless, NGC 4647 shows mild asymmetries in the
stellar and neutral gas distributions, and in the neutral gas
kinematics, suggesting an on-going interaction with NGC 4649
\citep{Young+06}.

Within the GC system, the hypothesis that blue GCs can trace the
interaction between the two galaxies more strongly than the red GCs
(e.g. \citealt{Forbes+97, Cortesi+13b}) is supported by the following
observations: i) the majority of GCs with redshifted velocities in NGC
4649 are blue, and are located on the NGC 4647 side
(Figs. \ref{fig:membership} and \ref{fig:fields}); and ii) the blue
GCs in the South and North West regions of NGC 1399 (close to the
nearby galaxies) have a broader line-of-sight velocity distribution
than the red GCs in the same region (Figs: \ref{fig:fields_sigma} and
\ref{fig:hist}).

\item {\it Kinematic disturbances from a satellite.} Kinematic disturbances
induced by the gravitational interaction with a satellite can
generate asymmetries in the stellar and ionized-gas velocity fields as
observed in interacting binary galaxies (e.g. \citealt{Borne+94}) and
galaxies in clusters (e.g. \citealt{Rubin+99}). However, stars, PNe
and GCs would  experience the same gravitational perturbation from
  the satellite, and therefore their velocity fields would be affected
  in the same way. Hence, this scenario would not cause differences in
  the kinematics between these tracers.

\item {\it Multiple mergers and accretion episodes.}  Massive galaxies in
dense environments are subject to many accretion events of different
kinds, i.e. major/minor, wet/dry mergers,  etc
(e.g. \citealt{DeLucia+07}).
 If the fractions of accreted GCs and stars/PNe differ for
  different accretion events, we may expect the kinematic signature of
  these events to be more visible in one system of tracers than in the
  other.

It is therefore interesting to compare the expected number of accreted
GCs and PNe for galaxies of different luminosity and type to see
  if there are any galaxies that would preferentially imprint the
  kinematic signature of the accretion in one system of tracers from
  the other.
In Figure \ref{fig:numbers} we compare the expected numbers of
accreted GCs and PNe as functions of the total magnitude and
morphological  type of galaxies. Despite the scatter, due
  to the low number statistics for dwarf galaxies and the different
  observational constraints for the PN and GC sample (which may add to an
  intrinsic scatter of the relation) there is an indication that low
luminosity and dwarf galaxies contribute more GCs than PNe. The
expected contributions from bright spiral galaxies are instead more
alike. Therefore, when a massive galaxy accretes many low luminosity
galaxies and dwarfs, it will preferentially accrete GCs. In this
  case, the signature of the merging events will be imprinted in the
  GCs kinematics more strongly than in the PNe and stars, therefore
  differences in the velocity fields of these tracers are to be
expected.

\begin{figure}
  \psfig{file=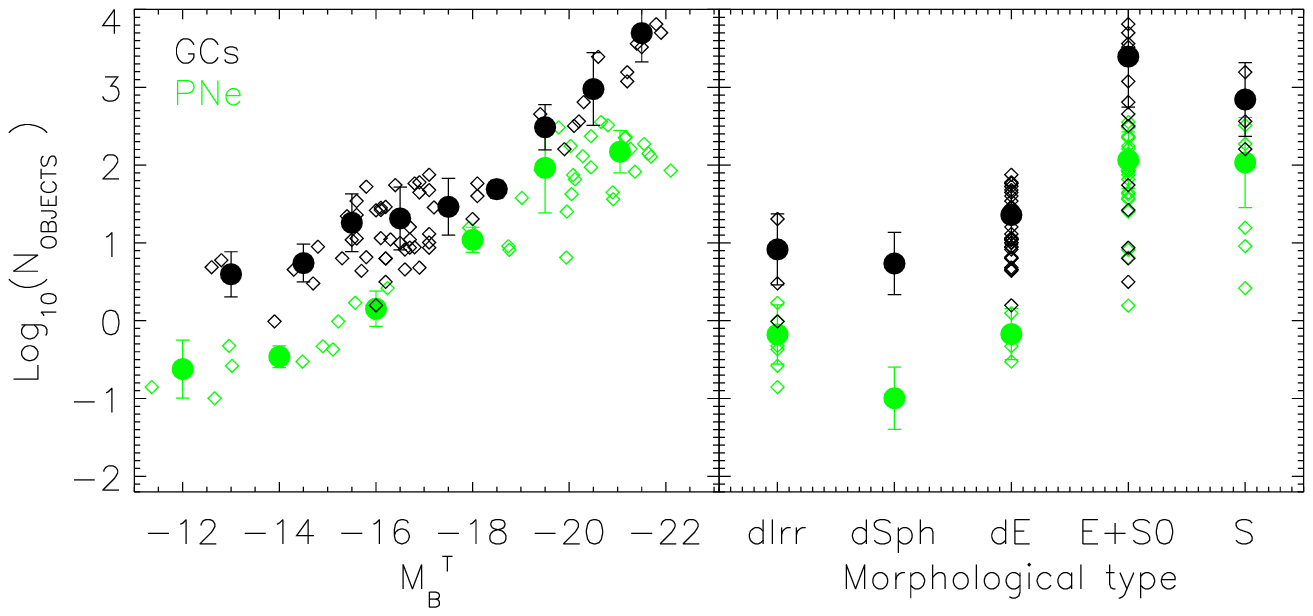,width=8.5cm,clip=,bb=56 350 430 532}
  \caption{Expected number of sources (GCS or PNe) as a function of
    galaxy absolute magnitude (left) and morphology (right). GCS data
    are from \citet{Harris+81}, \citet{BrodieStrader06}, and
    references therein. PN data are from \citet{Zijlstra+06},
    \citet{Buzzoni+06}, \citet{Coccato+09}, and references therein.
    The expected number of GCs and PNe are computed from: i) the
    specific frequencies of GCs and PNe, and ii) the total galaxies'
    luminosities in the $V$ and $B$ bands, for GCs and PNe
    respectively. The definitions of specific frequencies for GCs
      ($S_N$) and PNe ($\alpha_{\rm 1,B}$) are given by
      \citet{Harris+81} and \citet{Jacoby80}, respectively.}
\label{fig:numbers}
\end{figure}

This interpretation is consistent with the data of NGC 1399. In fact:
1) its globular cluster specific frequency ($S_N=12\pm4$,
\citealt{Harris+81}) is nearly twice the number expected for
  galaxies of the same total luminosity $M_B^T = -21.8$ ($S_{N,
  B=-21} \simeq 5$, see Figure \ref{fig:numbers}); and 2) its PN
specific frequency ($\alpha_{1.0, B} = 1.25 \pm 0.20 \cdot 10^{-9}
L_{\odot, B}$, \citealt{Buzzoni+06}) is nearly a third of that expected
for the galaxies of the same luminosity $M_B^T$ ($\alpha_{1.0,B=-22}
\simeq 4 \cdot 10^{-9} L_{\odot, B}$, see Figure
\ref{fig:numbers}). Therefore, these values are consistent with a
scenario where NGC 1399 accreted preferentially GCs from low
luminosity and dwarf galaxies.

 Similarly, the GC specific frequency in NGC 4649
($S_N=5.9\pm1.3$, \citealt{Harris+81}) is higher than expected from
galaxies of the same total luminosity $M_B^T = -21.4$ ($S_{N, B=-21}
\simeq 2$, from Figure \ref{fig:numbers}). Unfortunately, its PN
  specific frequency is not available in the literature.

 The situation of the merger remnant NGC 5128 is different because
  it is not in a dense environment. Recent accretions events, as
  suggested by the presence of substructures \citep{Woodley+11,
    Mouhcine+11, Crnojevic+13}, may come from the late infall of tidal
  tail material after the merger. In fact: 1) the GC specific
frequency ($S_N = 2.6 \pm 0.6$, \citealt{Harris+81}) is consistent
with the mean $S_N$ of galaxies of similar total $M_B^T=-19.97$
luminosity ($S_{N, B=-20} \simeq 3)$; and 2) its PN specific frequency
($\alpha_{1.0, B} = 12.5 \pm 0.6 \cdot 10^{-9} L_{\odot, B}$,
\citealt{Buzzoni+06}) is higher than that expected for galaxies of the
same luminosity ($\alpha_{1.0, B=-20} \simeq 1 \cdot 10^{-9} L_{\odot,
  B}$).

The difference in the PNe and GCs kinematics in NGC 5128 is similar to
the one recently found in the S0 galaxy NGC 2768 \citep{Forbes+12},
where the velocity PN field is indicative of a disk-like system,
whereas the GCs kinematics are more indicative of a spheroidal
component. It would be interesting to investigate the presence of a PN
sub-population associated with the spheroidal component in NGC 5128
with similar methodology as in \citet{Forbes+12} and \citet{Cortesi+13b}.

\end{itemize}

\section{Summary}
\label{sec:summary}

 In summary, we found that the kinematics of PNe, red GCs, and blue
  GCs are different in the three galaxies studied in this work. They
  contain different angular momentum vectors and their kinematics
  deviates from point-symmetry at large radii, indicating deviations
  from dynamical equilibrium. Tidal interactions, multi-epoch mass
  assembly and different specific frequencies depending on the
  progenitor galaxies can explain the occurence of different
  kinematics for different tracers.

It would be interesting to follow-up and compare the extended PNe and
GCs kinematics in other massive galaxies in dense environments, where
the evidence for accretion comes from independent arguments and
techniques, for example NGC 4889 \citep{Coccato+10a,
  Coccato+10b}, NGC 3311 \citep{Coccato+11, Arnaboldi+12}, M87
\citep{Mihos+05,Doherty+09,Romanowsky+12}, and NGC 1316 \citep{Richtler+12}.

\section*{Acknowledgments}
LC is supported by the European Community's Seventh Framework
Programme (/FP7/2007-2013/) under grant agreement No 229517.  We thank
K. Woodley and E. McNeil-Moylan for providing updated datasets for NGC
5128 and NGC 1399, C. Wegg and M. Lucio for useful discussion. We thank the
anonymous referee for useful comments that helped to improve the
paper.

\label{lastpage}
\bibliography{coccato12} 

%\end{document}
\appendix

\section{Determination of membership of tracers between galaxy and satellite in NGC 1399 and NGC 4649}

In the analysis of the kinematic data of NGC 1399 and 4649 we remove
the  tracers that belong to the satellites identified by
  probability $P \geq P_{\rm thr}$, where
$P_{\rm thr}=60\%$ (Section \ref{sec:halo}).

In this Section, we measure the dependency of our results on $P_{\rm
  thr}$, showing that there are no strong variations in the
interval $30\% < P_{\rm thr} < 90\%$. The is because the tracers with
uncertain membership are few, confirming the robustness of the object
separation procedure (see also \citealt{McNeil+10} for the stability
of the contaminants removal procedure in NGC 1399).

%\subsection{Rotation amplitude and direction for the PNe in NGC 1399}

\begin{figure}
  \psfig{file=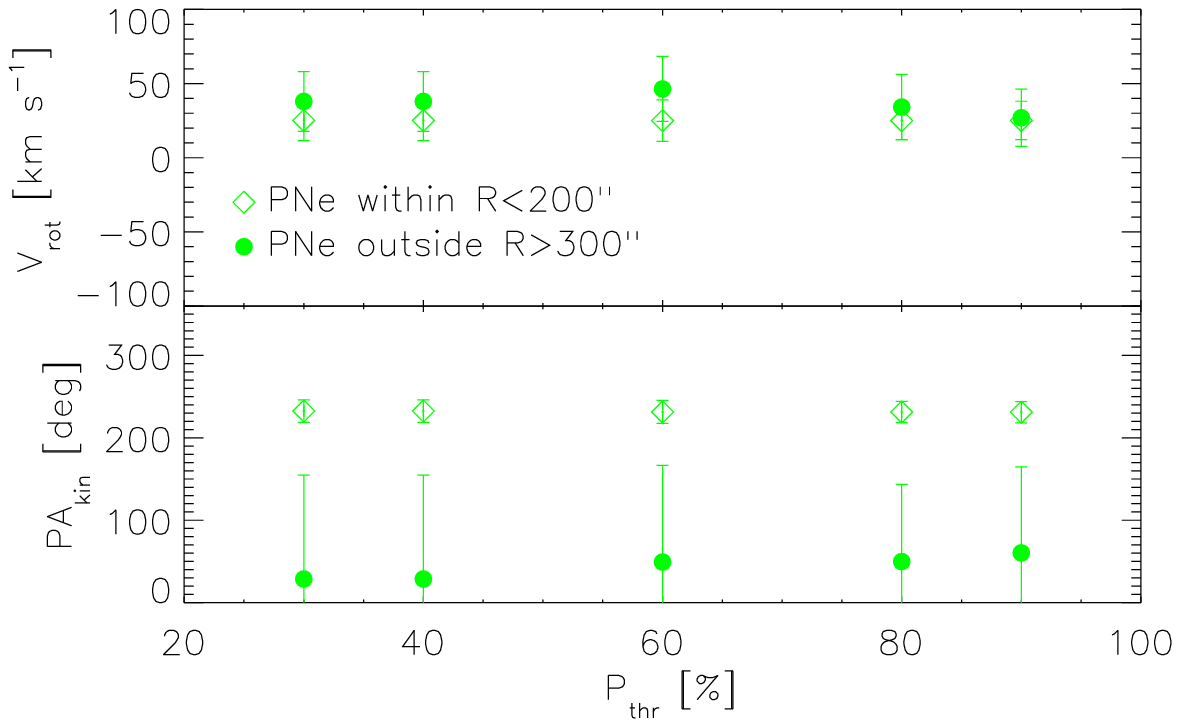,width=8. cm,clip=}
  \caption{Rotation amplitude (top panel) and direction (bottom panel)
    of PNe in the inner (open green dimanonds) and outer (filled green
    circles) regions in NGC 1399, as a function of different
    probability threshold $P_{\rm thr}$ used in the galaxy membership
    determination (see text for details).}
\label{fig:1399_pne}
\end{figure}

\begin{figure*}
  \begin{center}
\vbox{
  \hbox{
  \psfig{file=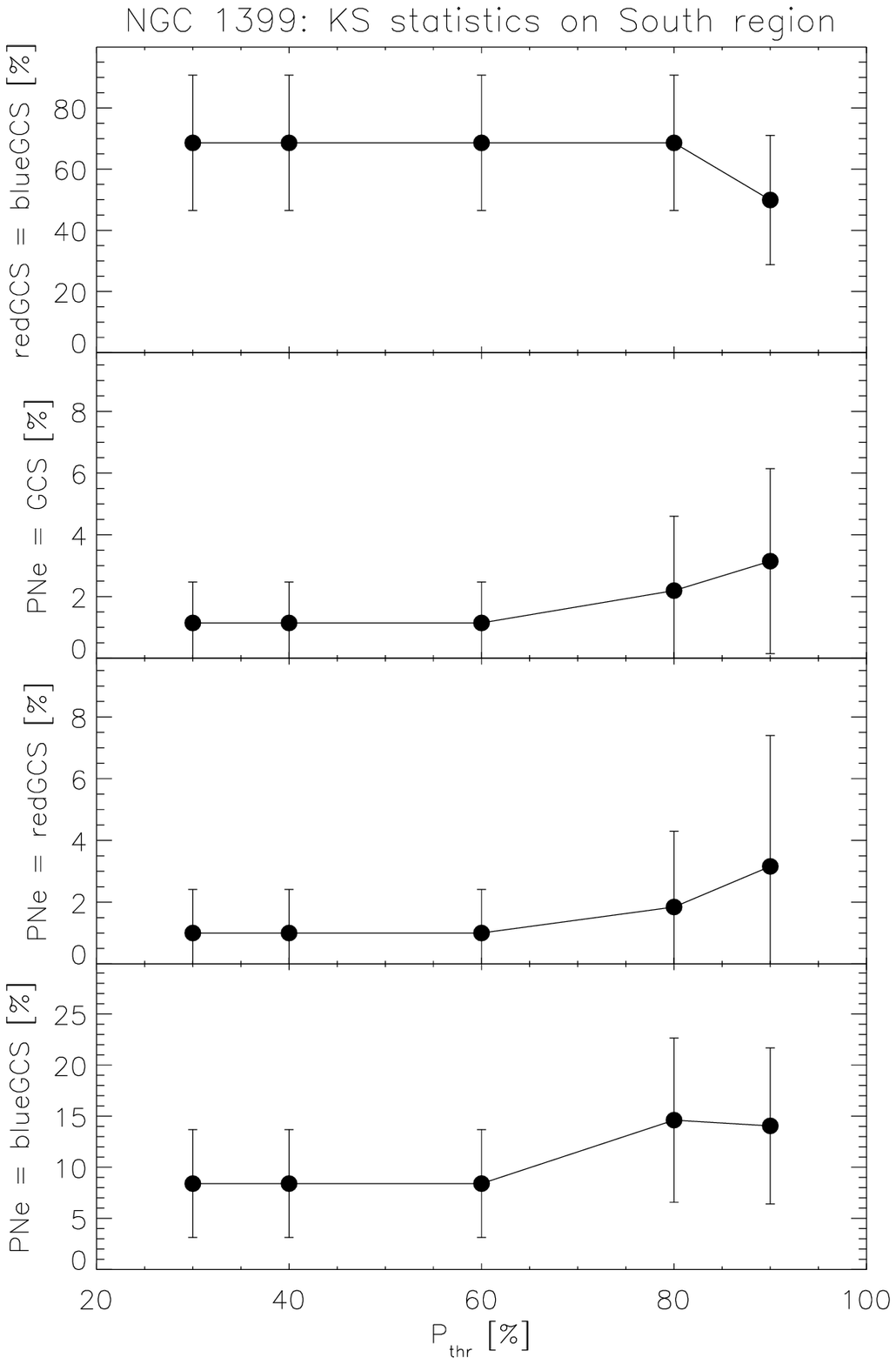,width=8.3cm,clip=}}
  \hbox{
  \psfig{file=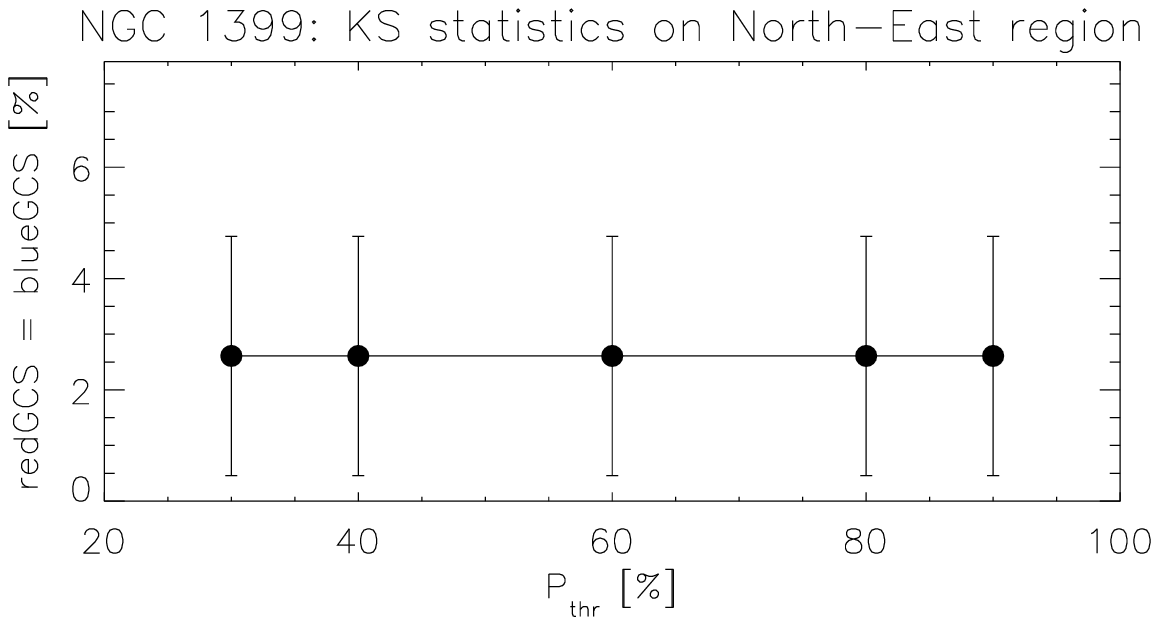,width=8.3cm,clip=}}
  \hbox{
  \psfig{file=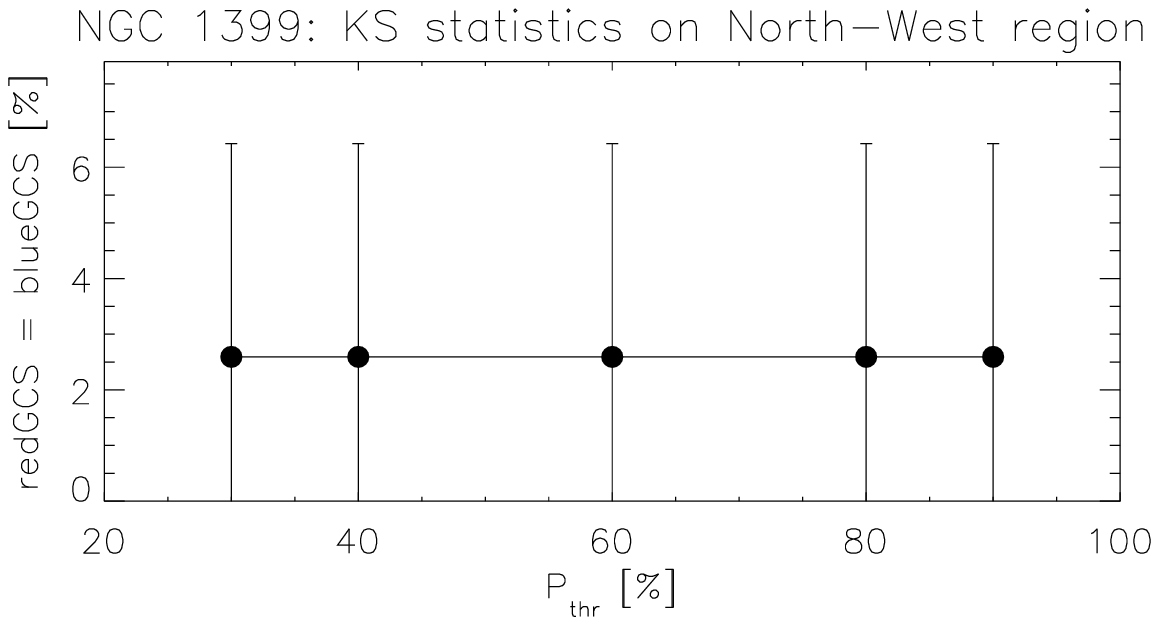,width=8.3cm,clip=}}
}
  \caption{Probability that the line-of-sight velocity distributions
    of two components are drawn from the same distribution, as a
    function of probability threshold $P_{\rm thr}$ for galaxy
    membership determination (see text for details).}
\label{fig:ks_threshold}
\end{center}
\end{figure*}

\begin{figure*}
  \psfig{file=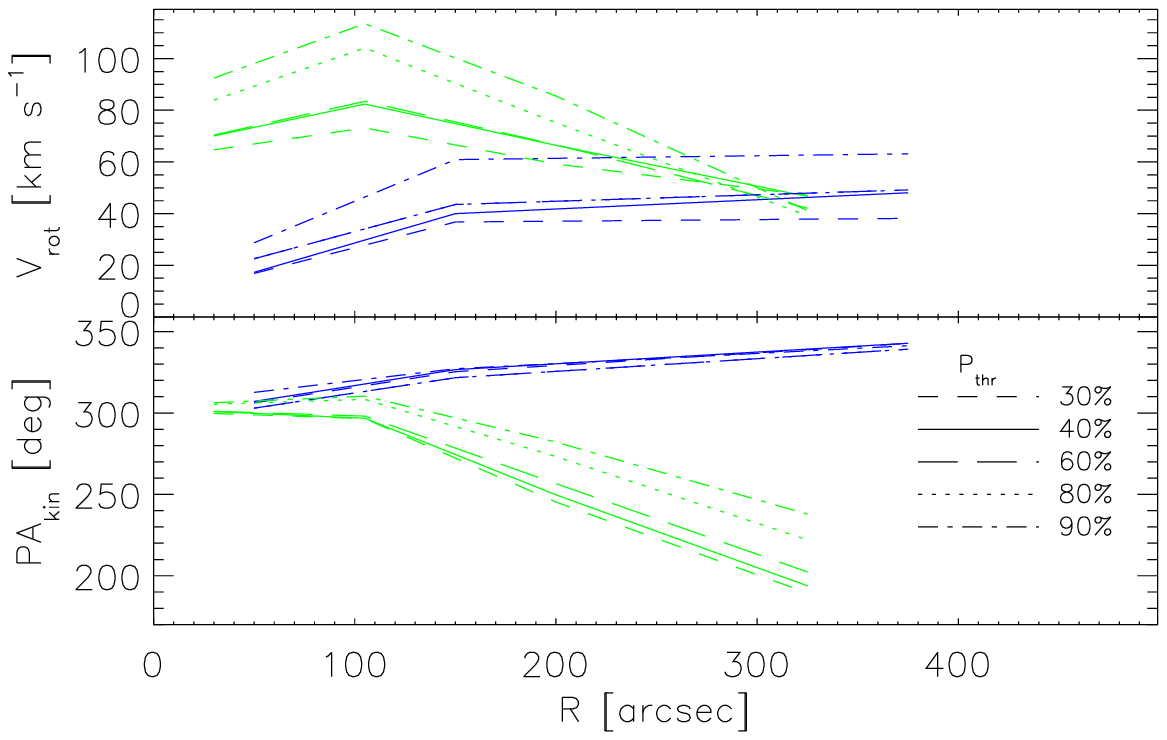,width=8.5cm,clip=}
  \caption{Rotation amplitude ($V_{\rm rot}$) and direction ($PA_{\rm
      kin}$) of PNe (green), and blue GCs (blue) systems in NGC 4649,
    as obtained from Equation \ref{eqn:kinem}. Different line styles
    represent the results adopting different $P_{\rm thr}$ (see text
    for details). Error bars have the same amplitude as those of
    Fig. \ref{fig:kinem4649}, but are not shown here for clarity.}
\label{fig:4649_threshold}
\end{figure*}

\begin{itemize}

\item In Section \ref{sec:1399} we claim that the PNe in the
  inner $R<200''$ and in the outer $R>300''$ in NGC 1399 rotate along
  opposite directions. In Figure \ref{fig:1399_pne} we show that the
  dependency of the amplitude and direction of rotation of these PNe
  on $P_{\rm thr}$ is neglegible.

%\subsection{The line-of-sight velocity distributions in NGC 1399}

\item In Fig. \ref{fig:hist} we show that PNe and GCs in three regions
  of the field of view have different velocity distributions. We
  quantified these differences with a Kolmogorov-Smirnov test,
  measuring the probability that PNe and GCs (and the red and blue GC
  sub-populations) are drawn from the same distribution.  In Figure
  \ref{fig:ks_threshold} we show that the Kolmogorov-Smirnov
  test does not vary more than a few per cent with $P_{\rm thr}$,
  confirming the robustness of our result.

%\subsection{Direction of rotation in NGC 4649}

\item In Sect. \ref{sec:4649} we claim that in the outer regions of
  NGC 4649 the PNe and blue GCS rotate along nearly opposite
  directions (Fig. \ref{fig:kinem4649}). In Figure
  \ref{fig:4649_threshold} we show that the direction of rotation is
  almost independent from $P_{\rm thr}$, confirming the robustness of
  the main result, i.e. the decoupling between the kinematic position
  angle of PNe and blue GCs. In contrast, the rotation amplitude of
  these tracers slightly depends on $P_{\rm thr}$: as $P_{\rm thr}$
  increases, more red-shifted objects are included in the analysis,
  increasing the best fit value of $V_{\rm rot}$.  In contrast to PNe
  and blue GCs, the red GC system is unaffected by $P_{\rm thr}$.

\end{itemize}

\newpage

\section{Mutual contamination between the red and blue GC sub-systems}

\begin{figure*}
\hbox{
  \psfig{file=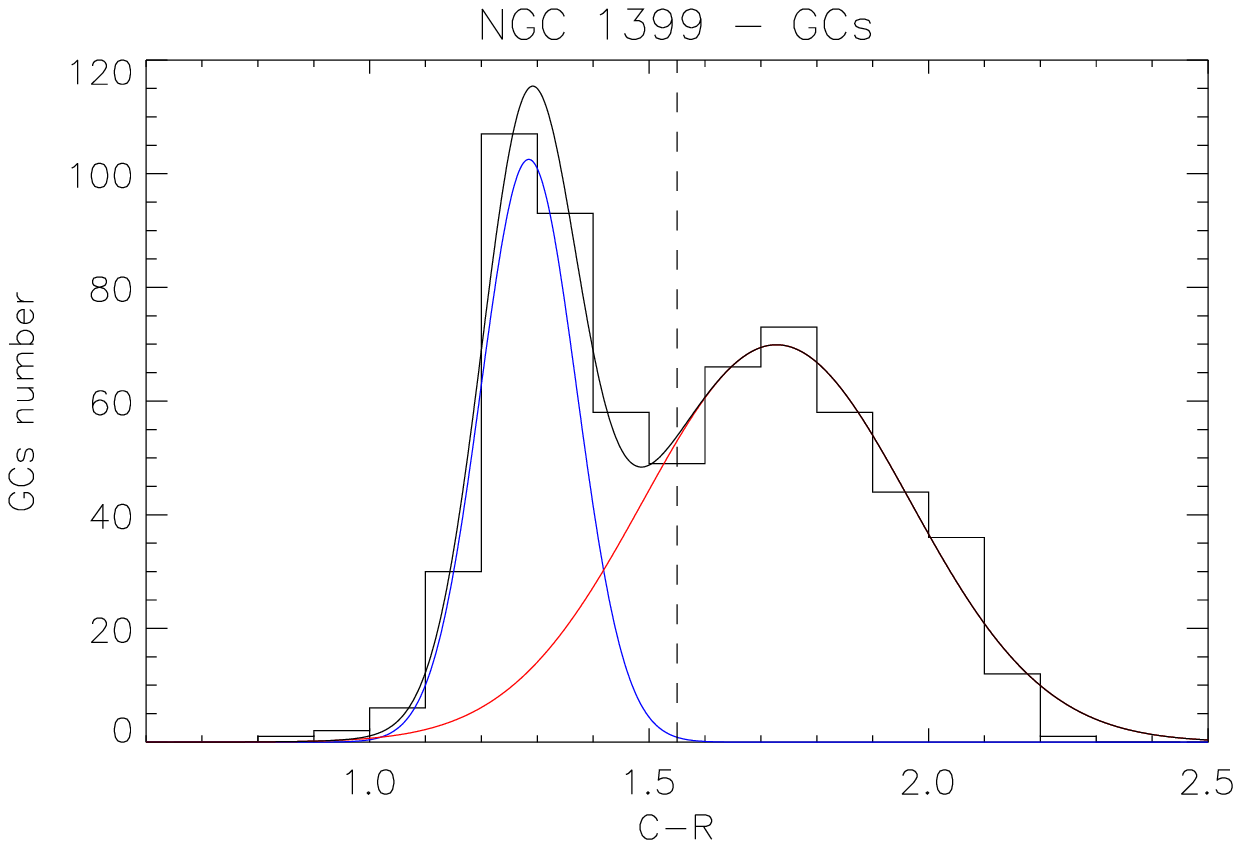,width=6cm,clip=}
  \psfig{file=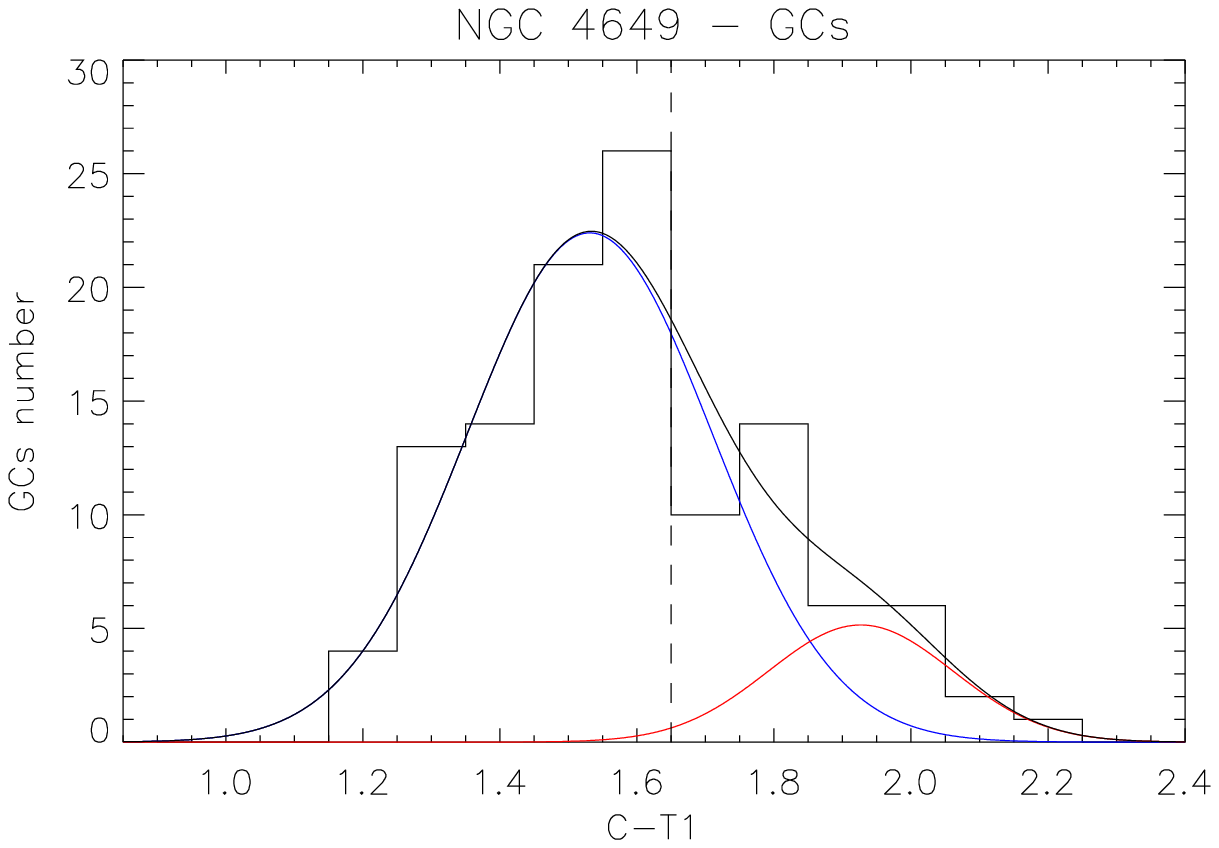,width=6cm,clip=}
  \psfig{file=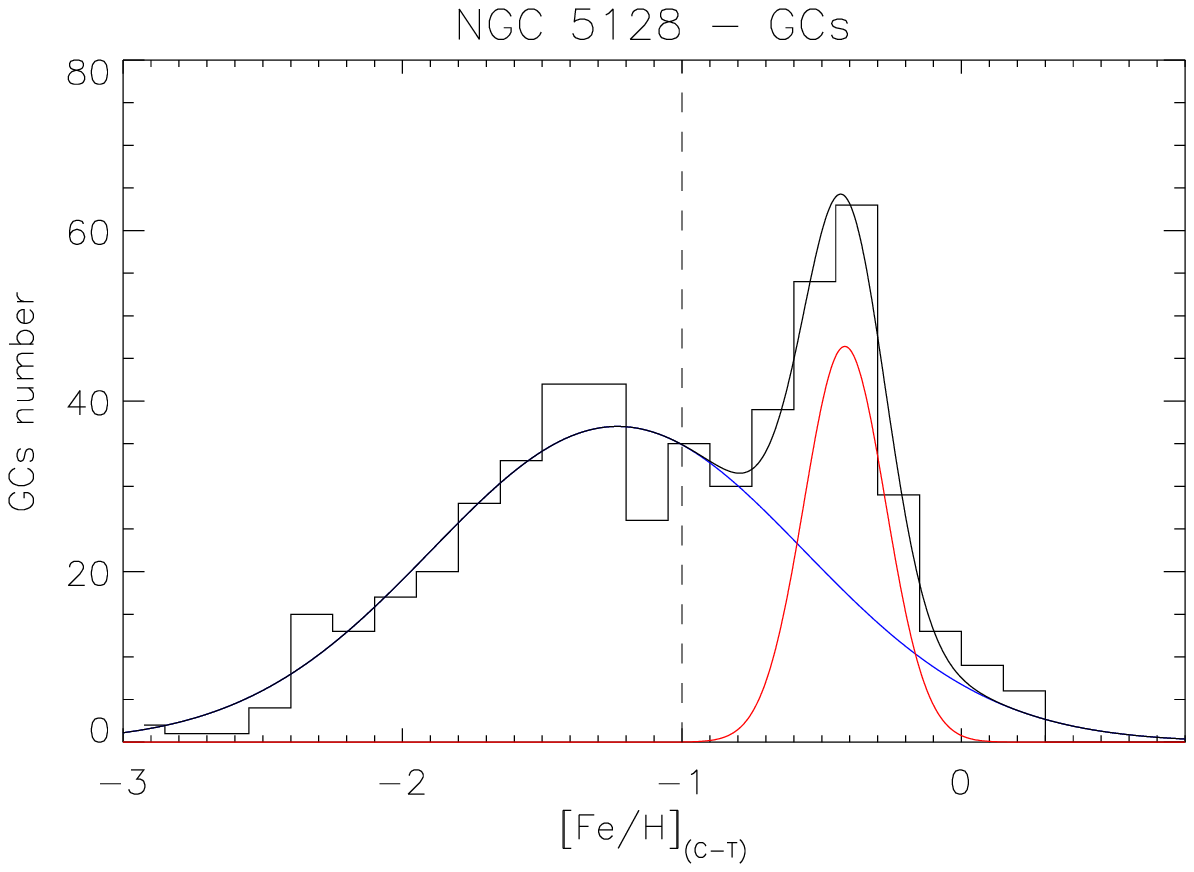,width=6cm,clip=}
}
\caption{Color distribution of the globular clusters used in our
    analysis: NGC 1399 (left panel), NGC 4649 (central panel), and NGC
    5128 (right panel). Red and Blue lines show the best-fit Gaussian
    functions describing the color distribution of the red and blue
    GC sub-systems, respectively. GCs classification is done from the
    color threshold (indicated by the vertical dashed line) defined in
    \citet[for NGC 1399]{Schuberth+10}, \citet[for NGC 4649]{Lee+08},
    and \citet[for NGC 5128]{Woodley+10b}.}
  \label{fig:bimodal}
\end{figure*}

We show in Fig. \ref{fig:fields} and \ref{fig:fields_sigma} and
discuss in Sect. \ref{sec:halo} that PNe and GCs have different
kinematic properties. We also divide the GCs into two sub-populations,
based on their color, and compare their kinematics with those derived
from PNe.  The color classification of GCs into red and blue is based
on the evidence that the color distribution is bimodal.  Nevertheless,
the two GC sub-populations overlap and a mutual contamination may be
present.  To quantify the amount of mutual contamination, we model the
GCs color distribution in our sample galaxies with two Gaussian
functions (Figure \ref{fig:bimodal}). Data and color separation
criteria are those defined in the papers that published the original
data.

The mutual contamination between the red and blue GC sub-systems is
quantified by measuring the area of the Gaussian functions that fit
the blue and red GCs ($N_{\rm BLUE}$ and $N_{\rm RED}$, respectively)
within the regions defined by the color thresholds.

In NGC 1399, the blue GC population is contaminated by the blue tail
of the red GC population ($N_{\rm RED} / N_{\rm BLUE} \simeq
0.4$). On the contrary, the red GC population is not contaminated by
blue GCs ($N_{\rm BLUE} / N_{\rm RED} \ll 10^{-2}$). Therefore, the
result that PNe and red-GCs have different line of sight velocity
distributions in the South region (Fig. \ref{fig:hist}) does not
change.

In NGC 4649 and NGC 5128, the red GC populations are strongly
contaminated by the red tail of the corresponding blue GC populations
($N_{\rm BLUE} / N_{\rm REF} \geq 1$), therefore their kinematic properties
  derived in Sections \ref{sec:4649} and \ref{sec:5128} depend on the
  adopted criteria for separation. However, the blue GC populations
are not contaminated by red GCs ($N_{\rm RED} / N_{\rm BLUE} \ll
10^{-2}$). Therefore, the result that PNe and blue-GCs in NGC 4649
rotate along different directions (Fig. \ref{fig:kinem4649}) does not
change.

%\end{document}
\end{document}